\documentclass{IEEEoj}
\IEEEoverridecommandlockouts
\usepackage[font=footnotesize]{caption}
\usepackage{cite}
\usepackage{fancyhdr,amsmath,amssymb,amsfonts}
\usepackage{algorithm2e}
\usepackage{algpseudocode}
\usepackage[pdftex]{graphicx}
\usepackage{epstopdf}
\usepackage{multirow}
\usepackage{textcomp}
\usepackage{xcolor}
\usepackage{spotcolor}
\usepackage[english]{babel}
\usepackage[printonlyused]{acronym}
\usepackage{authblk}
\usepackage{booktabs}
\usepackage{orcidlink}
\usepackage{subcaption}
\usepackage{adjustbox}
\usepackage{longtable}
\usepackage{multicol}
\usepackage{tabularx}
\usepackage{tikz}
\usepackage{float}
\usepackage{pgfplots}
\usepackage{pgfplotstable}
\pgfplotsset{compat=1.18}
\usetikzlibrary{shapes, arrows}
\usetikzlibrary{3d}
\usetikzlibrary{matrix}
\usetikzlibrary{decorations.pathreplacing}
\tikzset{
  dot/.style={fill=black,circle,minimum size=1pt,inner sep=0},
}
\SetKwComment{Comment}{/* }{ */}
\hypersetup{hidelinks}

\tikzstyle{block} = [rectangle, rounded corners, 
minimum width=2cm, 
minimum height=1cm,
text centered, 
draw=black, 
fill=blue!15]

\tikzstyle{arrow} = [thick,->,>=stealth]

\def\BibTeX{{\rm B\kern-.05em{\sc i\kern-.025em b}\kern-.08em
    T\kern-.1667em\lower.7ex\hbox{E}\kern-.125emX}}
\AtBeginDocument{\definecolor{ojcolor}{cmyk}{0.93,0.59,0.15,0.02}}
\def\OJlogo{\vspace{-14pt}\includegraphics[height=28pt]{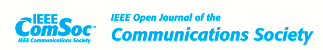}}

\begin{document}


\receiveddate{XX Month, 2024}
\reviseddate{XX Month, 2024}
\accepteddate{XX Month, 2024}
\publisheddate{XX Month, 2024}
\currentdate{XX Month, 2024}
\doiinfo{TBD}

\title{Interference analysis and modeling of Positioning Reference Signals in 5G NTN}

\author{
Alejandro Gonzalez-Garrido\authorrefmark{1},  \IEEEmembership{Student, IEEE}, \orcidlink{0000-0002-4695-8797},
Jorge Querol\authorrefmark{1}, \IEEEmembership{Member, IEEE},\orcidlink{0000-0002-8500-5534},
Henk Wymeersch\authorrefmark{2}, \IEEEmembership{Fellow, IEEE}, \orcidlink{0000-0002-1298-6159},
Symeon Chatzinotas\authorrefmark{1}, \IEEEmembership{Fellow, IEEE}, \orcidlink{0000-0001-5122-0001}
}

\affil{\authorrefmark{1}Interdisciplinary Centre for Security, Reliability and Trust. SnT. University of Luxembourg}
\affil{\authorrefmark{2}Department of Electrical Engineering, Chalmers University of Technology}

\authornote{This work has been submitted to the IEEE for possible publication. Copyright may be transferred without notice, after which this version may no longer be accessible}

\corresp{Corresponding author: Alejandro Gonzalez-Garrido, e-mail: alejandro.gonzalez@uni.lu}
\begin{abstract}

The integration of Positioning, Navigation, and Timing (PNT) services within the 5G non-terrestrial network (NTN) infrastructure is required to remove the requirement of a GNSS receiver in the user terminal. Using the positioning reference signal (PRS) in an NTN scenario presents significant challenges such as interference analysis from the transmission of several PRS. This study provides a stochastic model for the interference generated by PRS transmissions in a 5G NTN scenario. This model has been extracted empirically from a Monte Carlo simulator designed ad hoc for this purpose, showing the distribution that better fits the interference is a Generalized Extreme Value (GEV) distribution. The parameters of this distribution are also modeled in terms of the PRS configuration. Therefore, a designer can use this model to evaluate the probability of having certain levels of interference.  

\end{abstract}

\maketitle
\section{Introduction} \label{sec:Introduction}

\PARstart{O}{ne} key reason for extending \ac{5G} services to \ac{NTN} scenarios is the pursuit of global coverage for data services. From Release-17, \ac{NTN} \acp{UE} were mandated to incorporate a \ac{GNSS} receiver to access \ac{NTN} services \cite{el_jaafari_introduction_2023}. However, this requirement poses challenges for \ac{SNO}, as it limits their ability to offer their services in \ac{GNSS} denial areas. Furthermore, the power consumption of \ac{GNSS} receivers in \ac{IoT} devices could compromise their commercial viability. Consequently, developing a \ac{GNSS}-free \ac{UE} for \ac{NTN} operation is critical, motivating this study to explore offering \ac{PNT} services alongside data services through a unified \ac{NTN} infrastructure.
\ac{3GPP} in its release-16 of \ac{5G} has standardized positioning features to offer \ac{PNT} services \cite{noauthor_study_2019}. These features are distinguished by the deployment of various positioning techniques. Among the various positioning techniques in \ac{5G}, the use of a specific downlink signal, \ac{PRS}, is notable because it offers a wider bandwidth and higher carrier frequencies compared to previous generations, reaching up to 100~MHz in \ac{FR1} band and up to 400~MHz in \ac{FR2} band \cite{muursepp_performance_2021, panchetti_performance_2013}. However, the current definition of 5G \ac{PNT} services necessitates a network connection for the subscriber; in other words, 5G \ac{PNT} services are on-demand by the \ac{UE}, the core network, or a third party connected to the core network, in contrast to \ac{GNSS} that is a broadcast service. This architecture, inherited from \ac{LTE}, was initially designed for emergency call requirements as a terrestrial positioning system. In this framework, the network informs the \ac{UE} about the \ac{PRS} configuration; subsequently, the \ac{UE} acquires and relays the measurements back to the core network, which then performs position estimation. This approach poses scalability challenges in terms of the number of simultaneous users.
In addition, the \ac{5G} networks, from release-17 onward, include \ac{NTN} elements such as \ac{UAV}, \ac{HAPS}, and satellites. These components, increasingly emphasized by industry stakeholders, facilitate global communication capabilities \cite{rinaldi_non-terrestrial_2020}. Furthermore, in a \ac{NTN} satellite scenario each \ac{UE} require position information prior to its initial access. Therefore, current standardized positioning techniques are impractical due to the prerequisite of network connection for \ac{UE}.
Going beyond 5G, the 6G network is expected to establish a unified network entity, characterized by multiple connectivity layers designed to meet various device requirements in various scenarios \cite{trevlakis_localization_2023}. Therefore, the convergence of network's \ac{PNT} services with \ac{NTN} offers numerous advantages. These include the development of an autonomous integrated communication and navigation system under a unified network infrastructure, enhanced accuracy in \ac{PNT} solutions surpassing previous generations, global coverage that allows synchronized navigation and communication, increased resilience to positioning estimation, and the emergence of innovative services \cite{dureppagari_ntn-based_2023}. In addition, recent studies highlight the technological potential to achieve a truly integrated communication, location, and sensing system \cite{wei_5g_2023}. Furthermore, industry is also pushing forward, with the development of the Xona service, the LEO-PNT mission from ESA, and the Geesat constellation from ESA and China. However, these commercial development do not include a communication service.

\subsection{State of the Art}

Although the integration of communication and navigation systems within a unified network, as highlighted in the previous paragraph, holds the promise of revolutionizing \ac{PNT} solutions, it also introduces new challenges in signal interference management. The literature on interference in \ac{OFDM} systems, as illustrated \cite{martins_intersymbol_2019,cruz-roldan_intersymbol_2020,nemati_low_2018}, focuses mainly on terrestrial communication scenarios, focusing on \ac{ISI} issues in multipath environments. This existing research is pivotal for understanding interference dynamics; however, it mainly explores scenarios involving single transmitters. Furthermore, studies \cite{marijanovic_multiplexing_2020,kihero_inter-numerology_2019} delve into \ac{INI} modeling and improvement strategies, primarily in single-transmitter scenarios with \ac{ICI}, \ac{ISI}, and \ac{INI} as the main aggressors.
The complexity increases multifold when considering integrated systems, where multiple transmitters and receivers interact within a shared spectral environment. Similar approaches are already available in the literature for terrestrial networks, such as \cite{khan_novel_2019} where they proposed a \ac{FFR} scheme, or the concept of Network-MIMO, developed in \cite{venkatesan_network_2007} for indoor scenarios. However, none of them are intended for navigation systems. Therefore, this requires the development of advanced interference management techniques that can cater to the unique requirements of an integrated communication and navigation system, ensuring the reliability and accuracy essential for such a converged network infrastructure \cite{trabelsi_interference_2024}. This gap in existing research underscores the need for comprehensive studies that extend beyond traditional interference models to address the intricacies of integrated systems in future networks.
In a satellite scenario, the delays between the signals from different satellites (ms order) significantly exceed the length of the \ac{CP} of the waveform ($\mu$s order) \cite{zhen_energy-efficient_2021}. This leads to \ac{ISI} and \ac{ICI} at the receiver located in areas of overlapping inter-satellite beams. Assuming that all satellites are synchronized and that all satellites send positioning pilots synchronously in the same \ac{BWP}, two strategies are considered in the literature to address interference. The satellites apply a temporal guard band (called muting \cite{lin_positioning_2017}) long enough for the signal to reach each beam edge, or the satellites exploit the low probability of signal collision due to the large differential delay between satellites \cite{sixin_doppler_2022}
Despite significant advancements in interference analysis for \ac{5G} \ac{NTN} scenarios, a critical area remains underexplored: the aggregated interference effects caused by positioning signal transmissions, such as the \ac{PRS}, within an \ac{NTN} context. Current research does not thoroughly investigate the interference generated by differential propagation delays between satellites, which are considerably longer than the \ac{PRS} slot duration \cite{zhen_energy-efficient_2021}. It is essential for system designs to ensure that the interference levels between \ac{PRS} and data transmissions are minimized, allowing the receiver to accurately decode the data symbols and extract the observables of positioning.
This area of study is crucial for the development and optimization of \ac{PNT} services. By meticulously characterizing interference phenomena and understanding their impact on received signals, we can develop robust positioning algorithms and techniques that effectively mitigate interference effects. This will greatly improve the accuracy and reliability of positioning services, addressing the growing need for precision in contemporary applications.

\subsection{Paper Contributions}

In this paper, we evaluate the \ac{SINR} degradation due to multiplexing the broadcasting of \ac{PRS} among different satellite \ac{gNB}s in a \ac{BWP}. We assume the terrestrial multiplexing scheme for the \ac{PRS} signals \cite{dwivedi_positioning_2021} and translate it into a \ac{NTN} scenario.
Our proposal adopts an approach similar to that used by \ac{GNSS}, utilizing a dedicated \ac{BWP} for broadcasting the \ac{PRS}, termed \ac{BWPP}. Within the \ac{BWPP} the network operator broadcast the \ac{PRS} across all satellites. This requires a comprehensive analysis of the generated interference, ensuring that it remains sufficiently low for the receiver to decode the data symbols. Our study proposes an evaluation of interference in a system broadcasting positioning signals in a dedicated \ac{BWP}, therefore these signals are accessible to all users, similar to \ac{GNSS}. This approach enables \acp{UE} to initiate initial access to \ac{NTN} without requiring a GNSS receiver.
The primary objective of this research is to model the interference power received by a user terminal on the ground. The study focuses on the statistical characteristics of interference, critical for \ac{UE} positioning estimation. The contributions of this study include the following.
\begin{enumerate}
    \item Conducting a theoretical analysis of the received signal and the interference generated by the simultaneous reception of 4 \ac{PRS} at the receiver side. This interference is calculated at the output of the matched filter at the receiver.
    
    \item Develop a Monte Carlo simulator to evaluate the interference generated by the \ac{PRS}.The \ac{PRS} can be configured for a different number of symbols, combsize and transmitted power. It output the \ac{DDM} as a matched filter at the receiver for all received signals.

    \item Extract a novel stochastic model of the interference generated by the \ac{PRS}. This model is based on the configuration of the \ac{PRS} such as the transmitted power, the number of symbols, and the combsize used at the transmitter side. The model fitting is performed empirically with the results from the Monte Carlo simulator designed previously.
\end{enumerate}
The structure of this paper is as follows. Section \ref{sec:scenario} details the scenario and the channel model for \ac{LEO}. Section \ref{sec:Models} presents the signal models used in \ac{NTN}, the matched filter at the receiver, and the interference model. Section \ref{sec:InterferenceModelling} shows the Monte Carlo simulator developed and the stochastic model of interference extracted from the simulation results. Finally, Section \ref{sec:Conclusions} discusses the conclusions drawn from this study and suggests directions for future research.
\section{Scenario for 5G Satellite Positioning} \label{sec:scenario}

This section outlines the framework and scenario definition for a \ac{PNT} service provision via \ac{5G} \ac{NTN}. Involves a detailed examination of assumptions, simplifications, and the reasoning behind them. The proposed model requires that the signal from at least four distinct \ac{NTN} \ac{gNB}s similar to \ac{GNSS} reaches the user terminal.
In our study, we adopt an Earth-moving beam configuration, where the satellite beam moves along the satellite, as defined in 3GPP TR 38.821. without loss of generality, as the interference will be analyzed for single snapshot estimations. For multibeam satellites, the satellite implements precompensation at a reference ground point for each beam, effectively reducing the maximum delay/Doppler range experienced by the signal. Consequently, a single beam satellite represents a worst-case scenario from this perspective, which we focus on in our analysis.
Nowadays, beams overlapping can be fulfilled by massive constellations such as Starlink (our case assumes a single shell where all satellites are at the same altitude). There are some examples in the literature such as \cite{honnaiah_demand-driven_2023} on how to achieve this beam overlapping for a data service.
\subsection{Frequency reuse in a common 5G Resource Grid}
\begin{figure}
    \centering
    \begin{subfigure}{0.4\textwidth}
        \begin{tikzpicture}
            \tikzset{
                carrier/.style={fill=orange},
                bwp1/.style={fill=gray},
                freqarrow/.style={->, >=stealth},
                label/.style={text width=2.5cm, align=center}
            }
            \fill[carrier] (0,0) rectangle (5,1);
            \coordinate (A) at (0, 0); 
            \coordinate (B) at (-0.2, 0); 
            \coordinate (C) at (0, 1); 
            \fill[carrier] (A) -- (B) -- (C) -- cycle;
            \coordinate (A) at (5, 0); 
            \coordinate (B) at (5.2, 0); 
            \coordinate (C) at (5, 1); 
            \fill[carrier] (A) -- (B) -- (C) -- cycle;
            \node[label] at (2.5,1.3) {Operator Carrier};
            \fill[bwp1] (1,0) rectangle (4,0.9);
            \coordinate (A) at (1, 0); 
            \coordinate (B) at (0.8, 0); 
            \coordinate (C) at (1, 0.9); 
            \fill[bwp1] (A) -- (B) -- (C) -- cycle;
            \coordinate (A) at (4, 0); 
            \coordinate (B) at (4.2, 0); 
            \coordinate (C) at (4, 0.9); 
            \fill[bwp1] (A) -- (B) -- (C) -- cycle;
            \node[label] at (2.5,-0.3) {BWP1};
            \draw[freqarrow] (-0.5,0) -- (5.5,0) node[anchor=north west] {f};
        \end{tikzpicture}
        \caption{Single BWP within the operator carrier.}
    \end{subfigure}
    \hfill
    \begin{subfigure}{0.4\textwidth}
        \begin{tikzpicture}
            \tikzset{
                carrier/.style={fill=orange},
                bwp1/.style={fill=gray},
                bwp2/.style={fill=blue},
                bwp3/.style={cyan},
                bwp4/.style={magenta},
                freqarrow/.style={->, >=stealth},
                label/.style={text width=2.5cm, align=center}
            }
            \fill[carrier] (0,0) rectangle (5,1);
            \coordinate (A) at (0, 0); 
            \coordinate (B) at (-0.2, 0); 
            \coordinate (C) at (0, 1); 
            \fill[carrier] (A) -- (B) -- (C) -- cycle;
            \coordinate (A) at (5, 0); 
            \coordinate (B) at (5.2, 0); 
            \coordinate (C) at (5, 1); 
            \fill[carrier] (A) -- (B) -- (C) -- cycle;
            \node[label] at (2.5,1.3) {Operator Carrier};
            \draw[freqarrow] (-0.5,0) -- (5.5,0) node[anchor=north west] {f};
            \fill[bwp3] (2.5,0) rectangle (3.5,0.9);
            \coordinate (A) at (2.5, 0); 
            \coordinate (B) at (2.3, 0); 
            \coordinate (C) at (2.5, 0.9); 
            \fill[bwp3] (A) -- (B) -- (C) -- cycle;
            \coordinate (A) at (3.5, 0); 
            \coordinate (B) at (3.7, 0); 
            \coordinate (C) at (3.5, 0.9); 
            \fill[bwp3] (A) -- (B) -- (C) -- cycle;
            \node[label] at (3,-0.3) {BWP3};
            \fill[bwp2] (0.5,0) rectangle (2,0.9);
            \coordinate (A) at (0.5, 0); 
            \coordinate (B) at (0.3, 0); 
            \coordinate (C) at (0.5, 0.9); 
            \fill[bwp2] (A) -- (B) -- (C) -- cycle;
            \coordinate (A) at (2, 0); 
            \coordinate (B) at (2.2, 0); 
            \coordinate (C) at (2, 0.9); 
            \fill[bwp2] (A) -- (B) -- (C) -- cycle;
            \node[label] at (1.2,-0.3) {BWP2};
        \end{tikzpicture}
        \caption{Dual BWPs, each of them can transport different numerology and or services.}
    \end{subfigure}
    \hfill
    \begin{subfigure}{0.4\textwidth}
        \begin{tikzpicture}
            \tikzset{
                carrier/.style={fill=orange},
                bwp1/.style={fill=gray},
                bwp2/.style={fill=blue},
                bwp3/.style={cyan},
                bwp4/.style={magenta},
                freqarrow/.style={->, >=stealth},
                label/.style={text width=2.5cm, align=center}
            }
            \coordinate (O) at (0, 0);
            \coordinate (B) at (5, 0);
            \coordinate (C) at (0, 3);
            \draw[-latex] (O)  -- (B) node[at end, anchor=north]{$t$};
            \draw[-latex] (O)  -- (C) node[at end, anchor=south]{$f$};
            \fill[carrier] (0,0.1) rectangle (4.5,2.5);
            \fill[bwp1] (0,0.5) rectangle (1,2);
            \fill[bwp2] (1,1.25) rectangle (1.5,2.4);
            \fill[bwp3] (1,0.75) rectangle (1.5,1.25);
            \fill[bwp4] (1.5,1.2) rectangle (3.5,1.8);
        \end{tikzpicture}
        \caption{Dynamic allocation of BWP depending on the resource  needs.}
    \end{subfigure}
    \caption{Split of operator spectrum in different BWP in frequency and time.}
    \label{fig:BWP}
\end{figure}
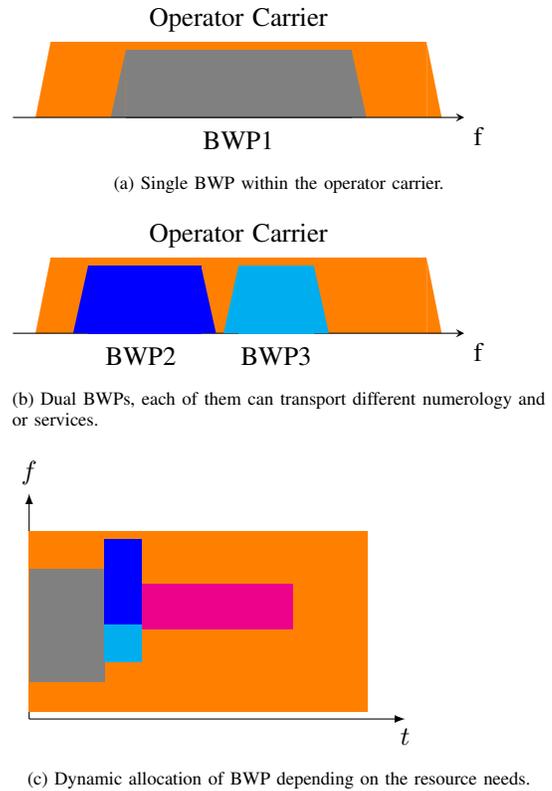

In a positioning system, the \ac{UE} has to receive several signals to obtain the observables. In the case of a 3D position using \ac{ToA} the \ac{UE} must receive the \ac{PRS} from at least four satellites to estimate its state as: position and clock bias, $[x,y,z,\delta(t)]$. A multiplexing scheme must be devised to address the high delays and Doppler shift characteristic of the \ac{NTN} channel, ensuring that the user can receive all four signals with minimal aggregate interference. In this regard, a 5G network operator can dynamically allocate its physical resources (time and frequency) based on the requirements of the user case. This dynamic allocation is referred to as \ac{BWP}, wherein the signals for different user profiles are partitioned in frequency or time, depending on the resources requested or available, as illustrated in Figure \ref{fig:BWP}.

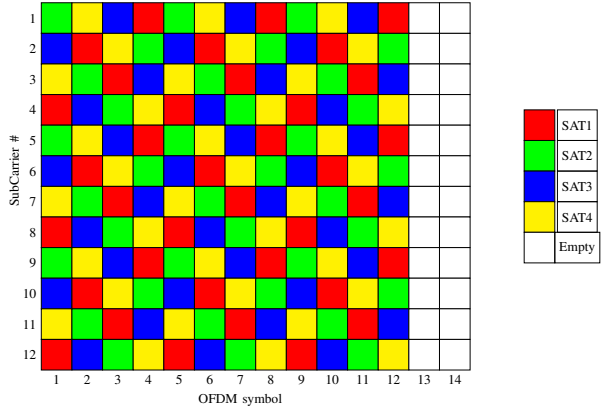
\begin{figure}
    \centering
    \begin{adjustbox}{width=0.45\textwidth,keepaspectratio}
        \begin{tikzpicture}
            \matrix (m) [matrix of nodes, nodes={draw, minimum size=8mm, anchor=center}, column sep=-\pgflinewidth, row sep=-\pgflinewidth, nodes in empty cells]{
            |[fill=green]| & |[fill=yellow]|& |[fill=blue]| & |[fill=red]| & |[fill=green]| & |[fill=yellow]|& |[fill=blue]| & |[fill=red]| & |[fill=green]| & |[fill=yellow]|& |[fill=blue]| & |[fill=red]| & |[fill=white]| & |[fill=white]| \\
            |[fill=blue]| & |[fill=red]|& |[fill=yellow]| & |[fill=green]| & |[fill=blue]| & |[fill=red]|& |[fill=yellow]| & |[fill=green]| & |[fill=blue]| & |[fill=red]|& |[fill=yellow]| & |[fill=green]| & |[fill=white]| & |[fill=white]| \\
            |[fill=yellow]| & |[fill=green]|& |[fill=red]| & |[fill=blue]| & |[fill=yellow]| & |[fill=green]|& |[fill=red]| & |[fill=blue]| & |[fill=yellow]| & |[fill=green]|& |[fill=red]| & |[fill=blue]| & |[fill=white]| & |[fill=white]| \\
            |[fill=red]| & |[fill=blue]|& |[fill=green]| & |[fill=yellow]| & |[fill=red]| & |[fill=blue]|& |[fill=green]| & |[fill=yellow]| & |[fill=red]| & |[fill=blue]|& |[fill=green]| & |[fill=yellow]| & |[fill=white]|& |[fill=white]| \\
            |[fill=green]| & |[fill=yellow]|& |[fill=blue]| & |[fill=red]| & |[fill=green]| & |[fill=yellow]|& |[fill=blue]| & |[fill=red]| & |[fill=green]| & |[fill=yellow]|& |[fill=blue]| & |[fill=red]| & |[fill=white]|& |[fill=white]| \\
            |[fill=blue]| & |[fill=red]|& |[fill=yellow]| & |[fill=green]| & |[fill=blue]| & |[fill=red]|& |[fill=yellow]| & |[fill=green]| & |[fill=blue]| & |[fill=red]|& |[fill=yellow]| & |[fill=green]| & |[fill=white]|& |[fill=white]| \\
            |[fill=yellow]| & |[fill=green]|& |[fill=red]| & |[fill=blue]| & |[fill=yellow]| & |[fill=green]|& |[fill=red]| & |[fill=blue]| & |[fill=yellow]| & |[fill=green]|& |[fill=red]| & |[fill=blue]| & |[fill=white]|& |[fill=white]| \\
            |[fill=red]| & |[fill=blue]|& |[fill=green]| & |[fill=yellow]| & |[fill=red]| & |[fill=blue]|& |[fill=green]| & |[fill=yellow]|& |[fill=red]| & |[fill=blue]|& |[fill=green]| & |[fill=yellow]| & |[fill=white]|& |[fill=white]| \\
            |[fill=green]| & |[fill=yellow]|& |[fill=blue]| & |[fill=red]| & |[fill=green]| & |[fill=yellow]|& |[fill=blue]| & |[fill=red]| & |[fill=green]| & |[fill=yellow]|& |[fill=blue]| & |[fill=red]| & |[fill=white]|& |[fill=white]| \\
            |[fill=blue]| & |[fill=red]|& |[fill=yellow]| & |[fill=green]| & |[fill=blue]| & |[fill=red]|& |[fill=yellow]| & |[fill=green]| & |[fill=blue]| & |[fill=red]|& |[fill=yellow]| & |[fill=green]| & |[fill=white]|& |[fill=white]| \\
            |[fill=yellow]| & |[fill=green]| & |[fill=red]| & |[fill=blue]| & |[fill=yellow]| & |[fill=green]| & |[fill=red]| & |[fill=blue]| & |[fill=yellow]| & |[fill=green]| & |[fill=red]| & |[fill=blue]| & |[fill=white]|& |[fill=white]| \\
            |[fill=red]| & |[fill=blue]|& |[fill=green]| & |[fill=yellow]| & |[fill=red]| & |[fill=blue]|& |[fill=green]| & |[fill=yellow]| & |[fill=red]| & |[fill=blue]|& |[fill=green]| & |[fill=yellow]| & |[fill=white]|& |[fill=white]| \\
            };
            \foreach \x in {1,...,14}
            \node[below] at (m-12-\x.south) {\x};
            \foreach \y in {1,...,12}
            \node[left] at (m-\y-1.west) {\y};
            \node[below=5mm] at (m-12-7.south) {OFDM symbol}; 
            \node[rotate=90,yshift=7mm] at (m-6-1.west) {SubCarrier \#}; 
            \matrix (m2) [matrix of nodes, nodes={draw, minimum size=8mm, anchor=center}, column sep=-\pgflinewidth, row sep=-\pgflinewidth, nodes in empty cells, yshift=0mm,xshift=80mm]{
            |[fill=red]| & SAT1 \\
            |[fill=green]| & SAT2 \\
            |[fill=blue]| & SAT3 \\
            |[fill=yellow]| & SAT4 \\
            |[fill=white]| & Empty \\
            };
        \end{tikzpicture}
    \end{adjustbox}
    \caption{5G PRS Transmitted Resource Grid example to multiplex 4 different satellites (each color is the PRS transmission from a different satellite). Size 1 Resource Block $\times$ 1 Slot and a Comb Size (frequency periodicity of the PRS per transmitter) of 4.}
    \label{fig:RG_12}
\end{figure}

In this study, we implement the \ac{PRS} multiplexing design used for terrestrial applications \cite{dwivedi_positioning_2021}, depicted in Figure \ref{fig:RG_12}. This design facilitates the transmission of multiple \ac{PRS} within a single \ac{OFDM} slot, whereby the empty \ac{RE} left by one transmitter, due to the steps of the subcarrier ("CombSize" parameter), is used by another \ac{gNB} for their \ac{PRS} allocation.

Unlike the terrestrial channel, the \ac{NTN} channel experiences larger differential delays and Doppler shifts. Our analysis is focused on the interference generated by different transmissions in this scenario, where each satellite's signal travels through a wireless channel that can be assumed to be independent and uncorrelated for each satellite. A challenge in this scenario is to minimize interference between transmissions from different satellites.
\subsection{Satellite Scenario}

\begin{figure}
    \centering
    \begin{adjustbox}{width=0.45\textwidth,keepaspectratio}
        \begin{tikzpicture}
            \def\LEOELRadius{4}
            \def\LEOELMINRadius{3.5}
            \draw (0,0) circle (\LEOELRadius cm);
            \draw[dashed] (0,0) circle (\LEOELMINRadius cm);

            \node[fill,circle,inner sep=1pt] (mainNode) at (0,0) {};
            \node[xshift=-5mm, yshift=-3mm] {Subsatellite Point (SSP)};
            \node[anchor=east] at (-0.707*\LEOELRadius,0.707*\LEOELRadius) {$\theta_{\text{MIN}}=0^\circ$};
            \draw[latex-latex] (0.707*\LEOELMINRadius,0.707*\LEOELMINRadius) -- (0.707*\LEOELRadius,0.707*\LEOELRadius) node[midway,right,xshift=5,yshift=5] {$\theta_{\text{MASK}}$};
            \draw[latex-latex] (0,0) -- (\LEOELRadius,0) node[midway, below] {$R_{\text{MAX}}$};
            \draw[latex-latex] (0,0) -- (-0.707*\LEOELMINRadius,0.707*\LEOELMINRadius) node[midway,left] {$\rho_{\text{MAX}}$};
        \end{tikzpicture}
    \end{adjustbox}
    \caption{Satellite maximum Field of View (FoV) (circle where the elevation angle mask is $0^\circ$). And satellite useful FoV (dashed circle) depends on the selected elevation angle mask $\theta_{\text{MASK}}$ and defines the maximum distance satellite-user $\rho_{\text{MAX}}$.}
    \label{fig:sat-DoC}
\end{figure}
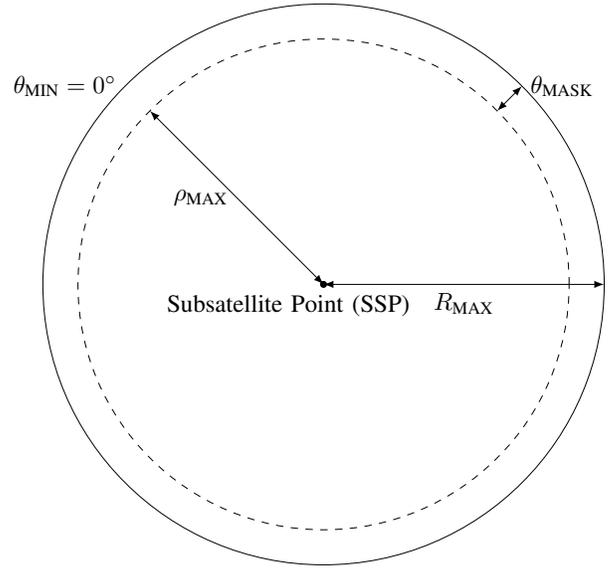

The initial step in evaluating a satellite system involves determining the service requirements, which, in turn, establishes a minimum \ac{SINR} at the perimeter of the service beam. Achieving this required \ac{SINR} primarily depends on mitigating the loss of the link, where the key factor is the distance between the satellite and the user at the edge of the beam marked as $\rho_{\text{MAX}}$ in Figure \ref{fig:sat-DoC}. This distance is essential to close the link budget.

Figure \ref{fig:sat-DoC} shows the \ac{FoV} of a satellite. Here, the furthest boundary represents the complete view of Earth from a satellite when a user is positioned at an elevation angle of $0^\circ$. It also delimits the maximum area of coverage by a radius $R_{\text{MAX}}$. At the center of this illustration lies the \ac{SSP}, which is the perpendicular projection of the satellite position on the Earth's surface. The dashed circle in Figure \ref{fig:sat-DoC} delimits the coverage area when users limit its operation to a certain elevation angle mask $\theta_{\text{MASK}}$, due to link budget constraints. The value of $\theta_{\text{MASK}}$ is crucial as it significantly impacts the maximum signal propagation time between the satellite and the ground station, as well as the maximal losses incurred due to \ac{FSPL}.
\begin{figure}
    \centering
    \begin{adjustbox}{width=\textwidth,height=0.35\textheight,keepaspectratio}
        \begin{tikzpicture}
            \def\hsat{4}
            \def\re{7}
            \coordinate (O) at (0,0);
            \coordinate (SSP) at (0,5);
            \coordinate (SAT) at (0,7);
            \coordinate (ELMIN) at (3.571,3.499);
            \coordinate (ELMASK) at (1.856,4.642);
            \coordinate (L_TAN) at (0,5.4);
            \coordinate (R_TAN) at (3.6,3.95);
            \draw (O)  -- (SSP) node[midway, anchor=east]{$r_E$};
            \draw (SSP) -- (SAT)[dotted];
            \begin{pgflowlevelscope}{\pgftransformscale{1}}
                \draw[decorate,decoration={brace,raise=1pt}] (SSP) -- (SAT) node[midway, anchor=east,xshift=-2pt]{$h_{\text{SAT}}$};
            \end{pgflowlevelscope}

            \draw (SAT) -- (ELMIN)[dotted];
            \draw (ELMIN) -- (O);
            \draw (SAT)[dashed] -- (ELMASK);
            \begin{pgflowlevelscope}{\pgftransformscale{1}}
                \draw[decorate,decoration={brace,raise=1pt}] (SAT) -- (ELMASK) node[midway, anchor=west,xshift=2pt,yshift=2pt]{$\rho_{\text{MAX}}$};
            \end{pgflowlevelscope}
            
            \draw (ELMASK)[dashed] -- (O);
            \draw (L_TAN) -- (R_TAN)node[at end, anchor=west]{Local Horizon (LH)};
            \node[fill,circle,inner sep=1pt] at (ELMASK) {};
            \draw[rotate=45] (ELMIN) rectangle +(-0.3,0.3);
            \draw[rotate=68.5] (ELMASK) rectangle +(-0.3,0.3);
            \node[fill,circle,inner sep=1pt,label=left:SSP] at (SSP) {};
            \node[fill,circle,inner sep=1pt,label=left:SAT] at (SAT) {};
            \node at (O) [below left] {$O$};
            \draw (ELMIN) arc(44.4:90:5);
            \draw[latex-latex] (0,1) arc(90:68.2:1) node[at start,anchor=east] {$\Theta$};
            \draw[latex-latex] (0,6) arc(270:308.2:1) node[midway,anchor=north, yshift=2] {$\varphi$ };
            
            \draw[latex-latex] (1,5) arc(150:124:1) node[midway,anchor=east,yshift=8,xshift=3] {\small{$\theta_{\text{MASK}}$ }};
        \end{tikzpicture}
    \end{adjustbox}
    \caption{Distances and angles within a satellite beam depends on the user local horizon (LH), and the altitude of the satellite $h_{\text{SAT}}$.}
    \label{fig:orbit_slant_range}
\end{figure}
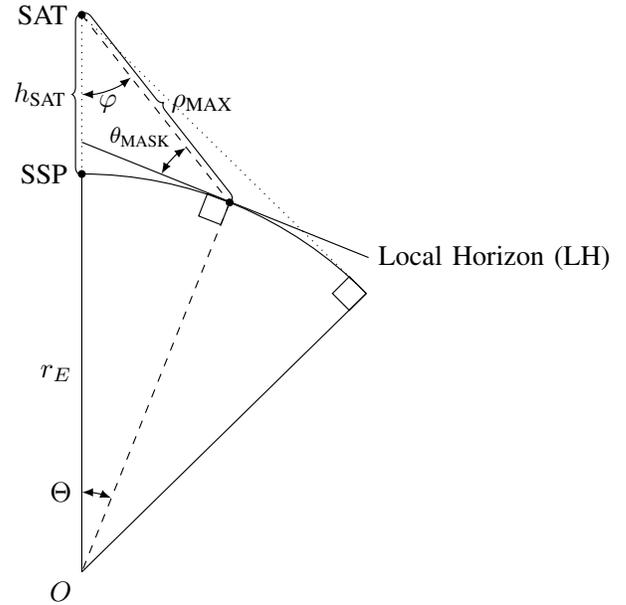

Figure \ref{fig:orbit_slant_range} shows as a perpendicular cross section of the plane illustrated in Figure \ref{fig:sat-DoC}, aiding in understanding the trigonometric calculations leading to \eqref{eq:rho_max}. This figure demonstrates the direct relationship between $\theta_{\text{MASK}}$ and $\rho_{\text{MAX}}$, in relation to the altitude of a satellite $h_{\text{SAT}}$ plus the radius of the Earth $r_{\text{E}}$.
\begin{equation}
    \rho_{\text{MAX}} = \left(r_E+h_{\text{SAT}}\right)\frac{\sin{\left(\frac{\pi}{2}+\Psi\right)}}{\sin{\left(\frac{\pi}{2}+\theta_{\text{MASK}}\right)}}
    \label{eq:rho_max}
\end{equation}
With the value of $\Psi$ defined as:
\begin{equation*}
    \Psi=-\theta_{\text{MASK}}-\arcsin{\left(\frac{r_E\sin{\left(\frac{\pi}{2}+\theta_{\text{MASK}}\right)}}{r_e+h_{\text{SAT}}}\right)}
\end{equation*}
Besides, the Local Horizon (LH) is the perpendicular plane to the Earth surface, at the user location, used to define the parameters of a satellite pass over this user.

\begin{figure}
    \centering
    \begin{adjustbox}{width=0.45\textwidth,keepaspectratio}
        \begin{tikzpicture}[every node/.style={font=\tiny}]
            \begin{scope}[canvas is zy plane at x=0]
                \draw[-latex] (-2,0) arc (0:43.5:-1.5cm and 2cm)node[midway, anchor=west]{$\theta_{\text{MAX}}$};
            \end{scope}
            \begin{scope}[plane x={(1,0)}, plane y={(0.3,1)},canvas is plane]
                \draw[dashed] (-2,0) arc (180:0:2cm and 2.5cm); 
                \draw (0,0) -- (0,2.5)node[midway, anchor=east, yshift=10,xshift=5]{$\rho_{\text{min}}$}; 
                \draw (0,0) -- (120:2.35)node[midway, anchor=east,yshift=8,xshift=-2]{$\rho(t)$};
                \draw[-latex] (-1.15,2.05) -- (90:3)node[at end, anchor=south]{$\mathbf{v}_{\text{SAT}}(t)$}; 
                \draw (0,0) -- (164:2)node[midway, anchor=south]{$\rho_{\text{MAX}}$}; 
                \fill (-1.25,2) rectangle ++(0.1,0.1)node[anchor=east]{Satellite};
            \end{scope}
            \begin{scope}[plane x={(1,-0.1)}, plane y={(0,1)},canvas is plane]
                \draw[-latex] (-1.8,0) arc (180:170:2cm and 2.5cm)node[midway, anchor=east]{$\theta_{\text{MASK}}$};
            \end{scope}
            \begin{scope}[canvas is zx plane at y=0]
                \draw (-2,-2) -- (2,-2) -- (2,2) -- (-2,2) -- cycle; 
                \draw[dotted] (0,0) -- (-2,0); 
                \draw[dotted] (0,0) -- (0,-2); 
                \node at (2,-2) [below right]{Local Horizon (LH) $[\phi,\lambda,h,t]$};
                \draw[-latex] (0,0) -- (0.5,0.5)node[at end, anchor=west]{$\mathbf{v}_{\text{UE}}(t)$}; 
                \fill (-0.05,-0.05) rectangle ++(0.1,0.1)node[anchor=north east]{UE};
            \end{scope}
        \end{tikzpicture}
    \end{adjustbox}
    \caption{Parameters involved in a single satellite pass.}
    \label{fig:LocalHorizon}
\end{figure}
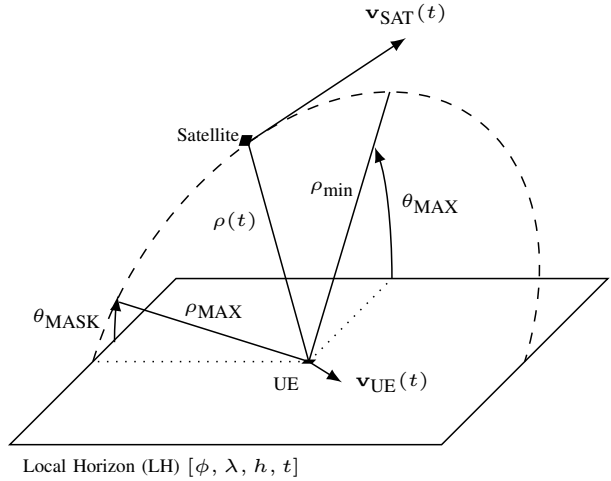

Figure \ref{fig:LocalHorizon} presents a three-dimensional representation of a single satellite pass, which is an expansion of the concept of LH from Figure \ref{fig:orbit_slant_range}. This depiction emphasizes that LH depends on the geographical coordinates of the user, defined by latitude $\phi$, longitude $\lambda$, altitude above mean sea level $h$, and time of observation $t$. This implies that a moving user's LH will change over time. However, for our analysis, as the user speed $\mathbf{v}_{\text{UE}}(t) \ll \mathbf{v}_{\text{SAT}}(t)$ is much smaller than the satellite speed, we could assume the user is static.
Moreover, Figure \ref{fig:LocalHorizon} highlights various parameters that play a critical role in understanding satellite dynamics from the perspective of a ground user, which are integral to the channel model. Among these parameters, $\theta_{\text{MAX}}$ stands out as particularly significant. It represents the maximum elevation angle that the satellite will attain during a specific pass over the user. This parameter is vital because it influences several other factors, such as the duration of the satellite pass and the minimum distance between the satellite and the user, represented by $\rho_{\text{min}}$. The range of $\theta_{\text{MAX}}$ is defined as being between $[\theta_{\text{MASK}},\pi/2]$. It should be noted that each satellite pass will have a unique value of $\theta_{\text{MAX}}$, which is determined by orbital dynamics and the specific location of the user.
\subsection{Wireless Channel Model}

We use a delay/Doppler spread representation of the wireless channel 
\cite{hong_delay-doppler_2022}, i.e., 
\begin{equation}
    \gamma_i(\upsilon,\tau) = \sqrt{L_i}~h_i\delta(\upsilon-\upsilon_i)\delta(\tau-\tau_i).
    \label{eq:channel_WSS}
\end{equation}
The channel representation in \eqref{eq:channel_WSS} depends on 4 parameters: the free space path loses $L_i=( {c}/{(4\pi f_{c}\rho_{i})})^2$, a random phase rotation $h_i$, a delay $\tau_i={\rho_i}/{c}$ and a Doppler shift defined as $\upsilon_{i}\triangleq-\frac{f_{c}}{c}\frac{d}{dt}\rho_{i}$. Where $\rho_{i}=\Vert\mathbf{r}_{\text{SATi}}-\mathbf{r}_{\text{UE}}\Vert$ is the distance between \ac{UE} and \textit{i}-th satellite, $\frac{d}{dt}\rho_{i}=\hat{\mathbf{u}}_{\text{SATi}}^T\mathbf{v}_{\text{SATi,UE}}$ is the relative speed between the \ac{UE} and \textit{i}-th satellite calculated as the projection of $\mathbf{v}_{\text{SATi,UE}}$ in $\hat{\mathbf{u}}_{\text{SATi}}$. Then, $\hat{\mathbf{u}}_{\text{SATi}} = \frac{\mathbf{r}_{\text{SATi}}-\mathbf{r}_{\text{UE}}}{\Vert\mathbf{r}_{\text{SATi}}-\mathbf{r}_{\text{UE}}\Vert}$ is a unitary vector that points from the user to the \textit{i}-th satellite, and $\mathbf{v}_{\text{SATi,UE}} = \left( \mathbf{v}_{\text{SATi}}-\mathbf{v}_{\text{UE}} \right)$ is the vector of velocity difference.
The channel \ac{FSPL} is modelled by $L_i$ assuming a unit gain on the TX and RX antennas. It depends on the carrier frequency $f_c$, and $\rho_i$. A more realistic \ac{NTN} channel has other losses, such as tropospheric effects (gas absorption, rain/cloud attenuation), antenna beam/polarisation misalignment, etc. These effects are assumed negligible as these attenuations compared to \ac{FSPL} are much lower for transmissions in L/S frequency bands.
The signal delay $\tau_i$ is also considered constant, as the change during a slot is negligible. A more accurate model would include an excess of ionospheric and tropospheric delay due to signal refraction. These effects have been extensively studied for \ac{GNSS} receivers and are modelled by the Klobuchar model \cite{klobuchar_ionospheric_1987} or the NeQuick model \cite{di_giovanni_analytical_1990, sanz_subirana_gnss_2013}. However, the inclusion of these models could obscure the analysis of this work that focuses on the effects on the performance of the satellite dynamics.
The model of $\upsilon$ shows that the measured Doppler is proportional to the relative speed of the satellite-user link in an ideal scenario, where its value is only affected by the dynamics of the satellite and the user. The channel model in \eqref{eq:channel_WSS} will serve as a baseline for the generation of a dataset published in \cite{qggt-xr49-24}. This dataset has several user positions, and for each location several satellite passes are computed with 1 second resolution. For each pass, the satellite position and velocity are stored.
Furthermore, we assume that the channel is \ac{WSS} for the duration of the slot (10~ms), thus, the values of $L_i$, $\tau_i$, and $\upsilon_i$ can be considered constant for the duration of the slot. This is a realistic assumption that does not compromise the results, as similar \ac{NTN} models use it \cite{baeza_overview_2022}.
\section{Transmitted and received signal model}\label{sec:Models}

In this section, we present the transmitted signal model and the theoretical framework for analyzing interference between satellites transmitting the \ac{PRS}.
Our focus is on analyzing the impact of the \ac{NTN} channel on transmissions from $S$ satellites, assuming \ac{LOS} and no multipath effects.

\subsection{Downlink Signal Model}

The \ac{5G} downlink signal model begins with the generation of the \ac{PRS} sequence for each \textit{i}-th satellite. This sequence is then mapped onto the \ac{RG} in accordance with 3GPP TS 38.211 Section 7.4.1.7.3 \cite{3gpp_nr_2024} as $\mathbf{A}_{i}\in\mathbb{C}^{M\times N_{\text{SCS}}}$ where $M$ are the \ac{OFDM} symbols and $N_{\text{SCS}}$ the total number of subcarriers of the Resource Grid. The \ac{CP-OFDM} modulation is applied by incorporating zero padding prior to the \ac{IFFT} operation as per 3GPP TS 38.211 Section 5.3.1 \cite{3gpp_nr_2024}. Following this, the \ac{CP} is appended by the transmitter. The transmitted signal from satellite $i$ is thus expressed in its complex baseband form, as described by 
\begin{equation}
    x_{i}(t)=\sqrt{P_{\text{TX}}}\sum_{m=0}^{M-1}\sum_{k=0}^{N_{\text{SCS}}-1}A_{i}[m,k]e^{j2\pi k \Delta ft}\text{rect}\left(\frac{t-mT_s}{T_s}\right),
    \label{eq:txsignal}
\end{equation}
where $T_s=T+T_{\text{CP}}$ indicates the symbol duration as the \ac{OFDM} symbol time $T$ plus the duration of the \ac{CP} $T_{\text{CP}}$, $\Delta f = {1}/{T}$ is the subcarrier spacing, and $\text{rect}(t/T_s)$ is the rectangular function. We assume the use of the rect pulse as it reduces the computational complexity and maintains the orthogonality between subcarriers, therefore the interference is due to the channel dynamics only.

The transmitted signal $x_{i}(t)$ has an average power level of $P_{\text{TX}}$. This power level is fixed at the satellite \ac{HPA} to guarantee minimum performance for beam edge users. This approach assumes, akin to \ac{GNSS}, a uniform \ac{EIRP} across the satellite beam.

\subsection{Received Signal Model}

The channel model outlined in \eqref{eq:channel_WSS} describes a channel between the \textit{i}-th satellite \ac{gNB} and the \ac{UE}. In a positioning system, the user typically receives all downlink signals within the same \ac{BWPP} spectrum. Thus, the received signal model is an aggregation of different \ac{NTN} signals, each affected by a distinct channel $\gamma_i$. The received signal is modeled by
\begin{equation}
    y(t) = \sqrt{L_{i}}\sum_{i = 0}^{\text{S}-1}e^{j2\pi \upsilon_i t+\varrho}x_{i}(t-\rho_i/c)+w(t),
    \label{eq:RX_signal}
\end{equation}
as the aggregation of the signal received by the $\text{S}$ satellites in \ac{LOS}. The model \eqref{eq:RX_signal} is essential for subsequent analyses, including \ac{SINR} evaluations and performance assessments of the delay estimator.

\subsection{Matched filter}

In the receiver architecture, the matched filter operation is based on the \ac{CAF}. This process involves correlating the received signal $y(t)$ with the different pilot signals shifted in frequency certain known value, analogous to the procedure employed by a \ac{GNSS} receiver in its acquisition phase. The definition \begin{equation}
    \chi_{xy}(\upsilon,\tau)=\int_{-\infty}^{+\infty}x(t)y^*(t-\tau)e^{j2\pi \upsilon t}dt.
    \label{eq:CAF_definition}
\end{equation}
illustrates the principle of detector operation, where the received signal is compared against the different local copies of the \ac{PRS}, one per satellite; therefore, the receiver will perform at least 4 different \ac{CAF} computations.

Substituting the received signal $y(t)$ into the \ac{CAF}, and following a similar analysis done in \cite{querol_snr_2016}, the matched filter output for the \textit{i}-th \ac{PRS} is given by 
\begin{align}
    \chi_{yx}^{(i)}(\upsilon,\tau)=&\sqrt{L_{i}}e^{j2\pi\left(\upsilon-\upsilon_{i}\right)\tau_{i}} \chi_{xx}^{(i)}\left(\upsilon-\upsilon_i,\tau-\tau_i\right)\notag \\
    + &\sum_{\substack{s \neq i \\ s \geq 0}}^{\text{Sat}-1}\sqrt{L_{s}}e^{j2\pi \left(\upsilon -\upsilon_s\right) \tau_s} \chi_{x_{s}x}^{(i)}(\upsilon-\upsilon_{s},\tau-\tau_{s}) \notag \\
    +&\chi_{wx}^{(i)}(\upsilon,\tau).\label{eq:detector_output}
\end{align}

\subsection{Post-matched filter Signal-to-Noise Plus Interference Ratio Analysis}

This subsection concludes the modeling discussion by presenting \ac{SINR} as a critical \ac{KPI} to analyze receiver performance. Assessing \ac{SINR} is paramount for the effective detection of the peak in the receiver's detector.
We know that the maximum value of the \ac{AF} is at the origin $\tau=0,\upsilon=0$, applying a variable change in \eqref{eq:detector_output} as $\tau'=\tau-\tau_i$ and $\upsilon'=\upsilon-\upsilon_i$ we displace the origin to the peak, then we evaluate the \ac{CAF} in relation to the difference in delay and Doppler of interference signals. For this we apply the following change of variable in \eqref{eq:detector_output} $\Delta\tau_s=\tau_i-\tau_s$ and $\Delta\upsilon_s=\upsilon_i-\upsilon_s$, yielding
\begin{align}
    \begin{aligned}
     &    \chi_{yx}^{(i)}(\tau',\upsilon')=\sqrt{L_{i}}e^{j2\pi\upsilon' \tau_i} \chi_{xx}^{(i)}\left(\upsilon',\tau'\right)\\
        &+ \sum_{\substack{s \neq i \\ s \geq 0}}^{S-1}\sqrt{L_{s}}e^{j2\pi \left(\upsilon' -\Delta\upsilon_s\right) \tau_s} \chi_{x_{s}x}^{(i)}(\upsilon-\Delta\upsilon_{s},\tau'-\Delta\tau_{s})\\
        &+ \chi_{wx}^{(i)}(\upsilon',\tau')
    \end{aligned}
    \label{eq:detector_output_change}
\end{align}
Therefore, we could find the contribution to the \ac{SNR} of the signal of interest $i$ from the other $s$ satellites. As mentioned above, the peak of the displaced \ac{CAF} correspond to $\upsilon'=0$ and $\tau'=0$, therefore, we define the \ac{SINR}  setting $\upsilon'=0$ and $\tau'=0$ in the \ac{CAF}:
\begin{align}
    \text{SINR}_{i} & = \frac{L_{i}P_{\text{TX}}}{\sum_{s \neq i}^{S-1}L_{s}|\chi_{x_{s}x}\left(\Delta\upsilon_{s},\Delta\tau_{s}\right)|^2 + |\chi_{wx}\left(\upsilon_i,\tau_i\right)|^2}
    \label{eq:SINR_definition}\\
   & =  \frac{1}{\sum_{s \neq i}^{S-1}\frac{L_{s}}{L_{i}}|\chi_{x_{s}x}\left(\Delta\upsilon_{s},\Delta\tau_{s}\right)|^2 + \frac{\sigma^2}{P_{\text{TX}}L_{i}}}
\end{align}
where we assumed the same transmission power across all satellites, denoted as $P_{\text{TX}}$, and the \ac{CAF} of the receiver noise as $|\chi_{wx}\left(\upsilon_i,\tau_i\right)|^2=\sigma^2$. This simplification incorporates the concept that the noise power is attenuated by the transmitted power normalised by the \ac{FSPL} at the \textit{i}-th satellite, represented as $P_{\text{TX}}L_{i}$.
Consequently, the interference contribution of the remaining satellites to the \textit{i}-th satellite \ac{SINR}, depends on the distance ratio $\rho$ as 
\begin{equation}
    I_i=\sum_{s \neq i}^{S-1}\left(\frac{\rho_{s}}{\rho_{i}}\right)^2 |\chi_{x_{s}x}\left(\Delta\tau_{s},\Delta\upsilon_{s}\right)|^2.
    \label{eq:interference}
\end{equation}
This assumption is valid under the approximation that all satellites transmit at the same wavelength, which is the characteristic scenario using a common \ac{BWP} for transmitting the \ac{PRS}. Moreover, interference is further influenced by differential delay, denoted as $\Delta\tau_{s}$, and differential Doppler shift, represented as $\Delta\upsilon_{s}$  between the satellite of interest $i$ and the interfere satellite $s$.
Deriving an analytical expression for $I_i$ is rendered infeasible due to the nature of the integral involved and the characteristics of the distance dynamics. Consequently, to analyse the effects of interference in a comprehensive way, we have implemented a Monte Carlo simulation and extract a \ac{PDF} for this interference.
\section{Interference modeling} \label{sec:InterferenceModelling}

This section describes the Monte Carlo simulator developed, the methodology to extract the interference model and the parameters models based on the \ac{PRS} waveform configuration.

\subsection{Monte Carlo simulator architecture}
\begin{figure}
    \centering
    \begin{adjustbox}{width=0.5\textwidth,height=\textheight,keepaspectratio}
        \begin{tikzpicture}[node distance = 2cm, auto]
        
            \node (sim_parameters) [block] {Simulation Parameters};
            \node (wav_gen) [block, below of=sim_parameters] {Waveform Generator};
            \node (channel) [block, below of=wav_gen] {Channel model};
            \node (acquisition) [block, below of=channel] {DDM Generation};
            \node (Performance) [block, below of=acquisition] {PDF modelling};
            \node (LEO_scenario) [block, right of=channel, xshift=2.5cm] {Dataset};
            
            \draw [arrow] (sim_parameters) -| node[anchor=south,xshift=-1cm]{user id, $t_{\text{pass}}$}(LEO_scenario);
            \draw [arrow] (sim_parameters) -- node[anchor=east] {$P_{\text{TX}},m,cs$}(wav_gen);
            
            \draw [arrow] (wav_gen) --node[anchor=east] {$x_{i}(t)$} (channel);
            \draw [arrow] (channel) -- node[anchor=east] {$y(t)$} (acquisition);
            \draw [arrow] (acquisition) -- node[anchor=east] {$I_{i}$} (Performance);

            \draw [arrow](LEO_scenario.west) |- node[midway,xshift=-0.75cm] {$L_{i},\tau_{i},\upsilon_{i}$} (channel.east);
            \draw [arrow] (LEO_scenario) |- node[anchor=south]{}(Performance);
            
            \draw [arrow] (wav_gen.west) -- ++(-1cm,0) |- (acquisition.west);
            
        \end{tikzpicture}
    \end{adjustbox}
    \caption{5G PRS LEO simulator architecture.}
    \label{fig:simulator}
\end{figure}
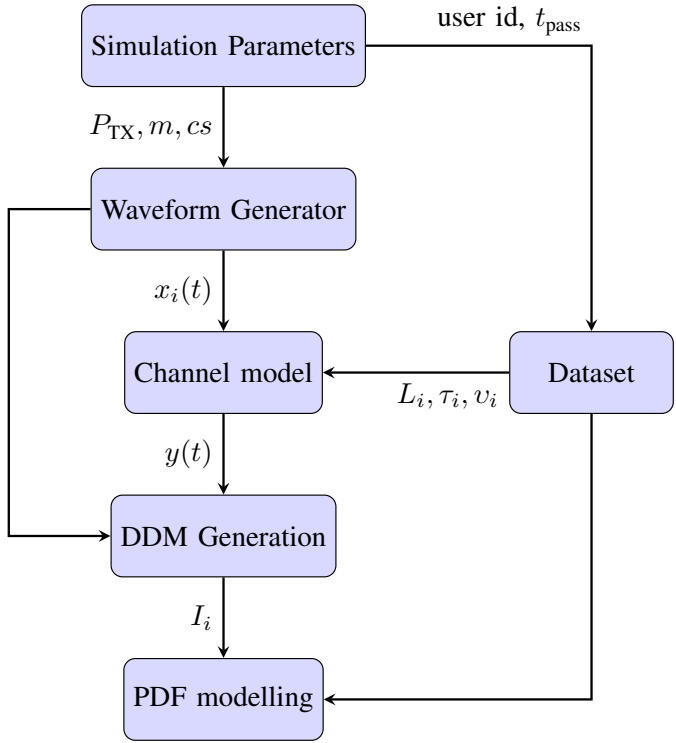

Figure \ref{fig:simulator} shows the simulator developed in order to extract the model for the \ac{PDF} of the interference seen in equation \eqref{eq:interference}. This simulator follows a Monte Carlo technique, evaluating the system for different user locations and signal configurations from the dataset \cite{qggt-xr49-24}

The assumption taken for the simulator is that all the satellites are synchronized and transmit the \ac{PRS} at the same time but in a different \ac{RG} arrangement, as seen previously in Figure \ref{fig:RG_12}.
This simulator starts with the definition of the simulation parameters, such as the signal configuration and the dataset \cite{qggt-xr49-24} for the satellite passes. Then it generates the waveforms requested. The next step is to apply to each waveform $i$ the corresponding channel, where the delay $\tau_{i}$, losses $L_{i}$ and Doppler $\upsilon_{i}$ are tightly coupled due to satellite movement. They also are dependant on the user's latitude, as some latitudes have more probability to have more satellites in view than others such as the poles, mid-latitudes or the Equator. Then the simulator performs the signal acquisition, by computing the \ac{DDM} of the received signal composed of the signal of interest and the other interference signals plus noise. Finally, from the \ac{DDM}, the simulator generates the interference power samples used to analyze its probabilistic behavior.

\begin{figure}
    \centering
    \includegraphics[width=0.45\textwidth]{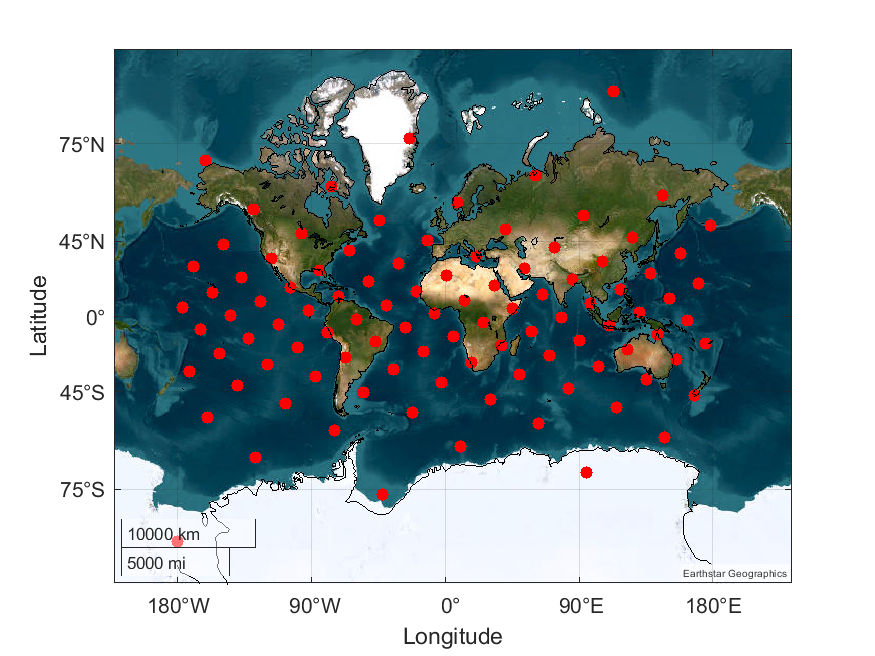}
    \caption{User's distribution of the dataset.}
    \label{fig:lattice}
\end{figure}

The dataset used is available in \cite{qggt-xr49-24}, it has 100 users uniformly spread on Earth's surface (using a Fibonacci lattice). For each user location, 10 min of Starlink satellite passes are stored with 1~s resolution. Figure \ref{fig:lattice} shows the location of the users on Earth from the dataset used to compute the satellite passes.

The Monte Carlo simulator performs 1000 iterations per user location. For each iteration, the simulator took a random time to obtain the delay, Doppler, and distance between the user and 4 satellites that have an elevation angle greater than a selected mask $\theta_{\text{MASK}}\leq\theta_{i}$.

\begin{table}
    \centering
    \begin{tabular}{lll}
        \hline
        \textbf{Description}   & \textbf{Symbol} & \textbf{Value} \\ \hline
        Number of satellite in LOS & $S$  & 4 \\ \hline
        Maximum signal bandwidth & $BW_{\text{MAX}}$ &  8.64~MHz \\ \hline
        Satellite's constellation &   & \begin{tabular}{@{}c@{}}Starlink. Inclination \\ of 53 deg/554km \end{tabular}  \\ \hline
        Transmitter power    & $P_{\text{TX}}$ & 1,10,20,30 dBW \\ \hline
        Carrier frequency & $f_c$ & n256 (2.2~GHz) \\ \hline
        Number of MC iterations & $N_{\text{index}}$ & 1000 \\ \hline
        Doppler Max value & $\pm \upsilon_{\text{MAX}}$ & 40~kHz \\ \hline
        Doppler resolution    & $\upsilon_{\text{step}}$ & 500~Hz \\ \hline
    \end{tabular}
    \vspace{3mm}
    \caption{Scenario details}
    \label{tab:framework}
\end{table}

Table \ref{tab:framework} enumerates the parameters relevant to the scenario delineated in Section \ref{sec:scenario}. In this scenario, the number of concurrent satellites in \ac{LOS} is set to four, which represents the minimum required for a 3D position estimation. The bandwidth is designated as the minimum permissible for the transmission of the \ac{PRS}. Similarly, the carrier frequency is chosen as the highest allowed within the n256 band.
\begin{table}
    \centering
    \begin{tabular}{lll}
        \hline
        \textbf{Description}   & \textbf{Symbol} & \textbf{Value} \\ \hline
        Number of Symbols & $m$  & 1 to 12 \\ \hline
        Number of Subcarriers & $N_{SC}$ &  288 \\ \hline
        Subcarrier Spacing &  $\Delta f$ &  30~kHz \\ \hline
        Comb Size & $cs$ & 4, 6 and 12 \\ \hline
    \end{tabular}
    \vspace{3mm}
    \caption{PRS generation details.}
    \label{tab:PRS_details}
\end{table}

Table \ref{tab:PRS_details} shows the different values of the parameters used to generate the \ac{PRS} used in the simulation. We have made a comparison using different number of \ac{OFDM} symbols, different values of CombSize, and different transmission powers.

\subsection{Interference distribution model extraction}

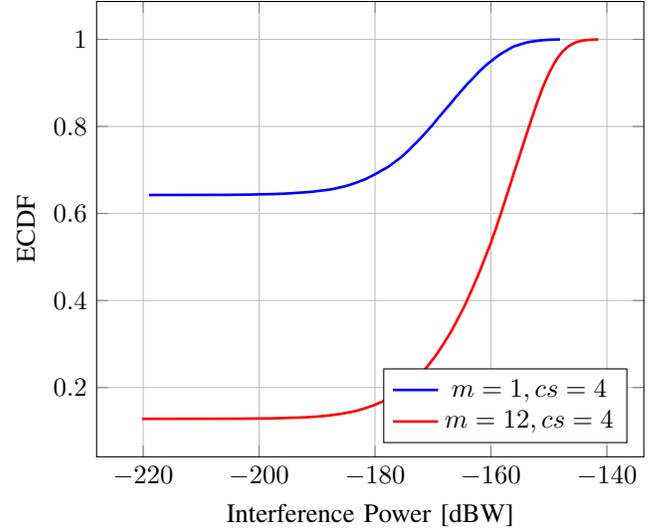
\begin{figure}
    \centering
    \begin{tikzpicture}
        \begin{axis}[
            xlabel={Interference Power [dBW]},
            ylabel={ECDF},
            legend pos={south east},
            grid=both,
            width=0.5\textwidth,
        ]

        \addplot[
            color=blue,
            solid,
            line width=1pt,
        ] table [x=x1, y=y1, col sep=space] {ECDF_S1.dat};
        \addlegendentry{$m=1,cs=4$}

        \addplot[
             color=red,
             solid,
             line width=1pt,
        ] table [x=x1, y=y1, col sep=space] {ECDF_S12.dat};
        \addlegendentry{$m=12,cs=4$}

        \end{axis}
    \end{tikzpicture}
    \caption{\ac{ECDF} comparison between using 1 or 12 symbol OFDM and a CombSize of 4.}
    \label{fig:ecdf_example}
\end{figure}

As seen in the previous Section \ref{sec:Models} the statistical interference model in \eqref{eq:interference} used in the \ac{SINR} analysis cannot be expressed in a closed form. Therefore, here we present the methodology for using the Monte Carlo approach to extract the distribution of the interference created by the PRS.
Figure \ref{fig:ecdf_example} illustrates a comparison between two examples \ac{ECDF} for two waveform configurations. One is using one symbol \ac{OFDM} and the other uses 12 symbols, both with a combsize of 4 and a transmission power of $30$~dBW. The plot reveals that with an $80\%$ probability, using one symbol, the interference power will be at least $-170$~dBW, while using 12 symbols, these values increase up to $-150$~dBW.

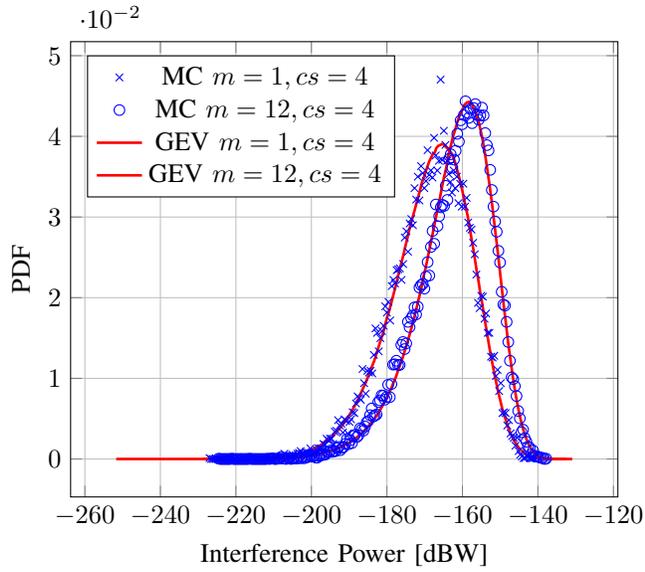
\begin{figure}
    \centering
    \begin{tikzpicture}
        \begin{axis}[
            xlabel={Interference Power [dBW]},
            ylabel={PDF},
            legend pos=north west,
            grid=both,
            width=0.5\textwidth,
        ]

        \addplot[
            color=blue,
            only marks,
            mark=x,
            mark options={},
        ] table [x=bin_centers1, y=pdf_counts1, col sep=space] {PDF_MC_dist.dat};
        \addlegendentry{MC $m=1,cs=4$}

        \addplot[
            color=blue,
            only marks,
            mark=o,
            mark options={},
        ] table [x=bin_centers12, y=pdf_counts12, col sep=space] {PDF_MC_dist.dat};
        \addlegendentry{MC $m=12,cs=4$}

        \addplot[
            color=red,
            solid,
            line width=1pt,
        ] table [x=bin_centers_dist1, y=pdf_counts_dist1, col sep=space] {PDF_MC_dist.dat};
        \addlegendentry{\ac{GEV} $m=1,cs=4$}

        \addplot[
            color=red,
            solid,
            line width=1pt,
        ] table [x=bin_centers_dist12, y=pdf_counts_dist12, col sep=space] {PDF_MC_dist.dat};
        \addlegendentry{\ac{GEV} $m=12,cs=4$}

        \end{axis}
    \end{tikzpicture}
    \caption{Example of a \ac{PDF} of the Interference power to compare the results from the Monte Carlo and the \ac{GEV} distribution.}
    \label{fig:PDF_Interference}
\end{figure}

To find the probability distribution that best describes this behavior, a \ac{KS} metric is used to evaluate the fitness of the measurement samples to a tested distribution. The \ac{ECDF} obtained from the interference power samples in the simulator are compared to different distributions, and among them the one with the lowest \ac{KS} metric is selected to model it. This \ac{KS} metric is the largest distance between the \ac{ECDF} of the samples and the \ac{CDF} of the different distributions tested. \begin{equation}
    D_{n,m}=\sup_{x} |F_{n}(x) - G_{m}(x) |
\end{equation}
Where $F_{n}(x)$ and $G_{m}(x)$ are the \ac{ECDF} of the samples and the \ac{CDF} of the distribution under test. To this test, we have added the p-value of the \ac{KS} test as it indicates the probability of obtaining test results at least as extreme as the observed results, under the assumption that the null hypothesis (the sample follows the reference distribution) is true. A higher p-value indicates a better fit to the distribution. 
\begin{table}
    \centering
    \begin{tabular}{|c|c|c|}
        \hline
        Distribution                & \ac{KS}& p-test   \\ \hline
        Normal                      & 0.0382 & 0        \\ \hline
        LogNormal                   & 0.0256 & 0        \\ \hline
        Gamma                       & 0.0298 & 0        \\ \hline
        Rayleigh                    & 0.5295 & 0        \\ \hline
        Rician                      & 0.0382 & 0        \\ \hline
        \ac{GEV}  & 0.0142 & 0.368    \\ \hline
    \end{tabular}
    \caption{\ac{KS} test result for the different distribution tested.}
    \label{tab:KS_test}
\end{table}
Table \ref{tab:KS_test} shows the \ac{KS} fitness test and p-value for the different distribution used to test. The distribution that best fits the measurements corresponds to the lower \ac{KS} statistic of 0.0142 and the highest p-value of 0.3688 corresponds to a \ac{GEV} distribution. The \ac{CDF} expression for this \ac{GEV} distribution is described as
\begin{equation}
    F_{\text{Interference}}(x) = \exp{\left(-\left[1+k\left(\frac{x-\mu}{\sigma}\right) 
\right]^{-1/k}\right)}.
    \label{eq:CDF_completa}
\end{equation}
This \ac{GEV} distribution is defined by 3 parameters: $k$ or shape, $\sigma$ or scale, and $\mu$ or location.
Figure \ref{fig:PDF_Interference} shows in blue the empirical \ac{PDF} for two different waveform configurations (using 1 and 12 OFDM symbols) and with different colors, in red, the fitted \ac{PDF} using a \ac{GEV} model.
Once the distribution that best fits the samples is found, we derive an empirical model to link the \ac{PRS} configuration settings with the distribution parameters. Each parameter of the \ac{GEV}, $[k,\mu,\sigma]$, depends on the \ac{PRS} configuration such as the number of symbols $m$, combsize $cs$, and transmitted power $P_{\text{TX}}$.
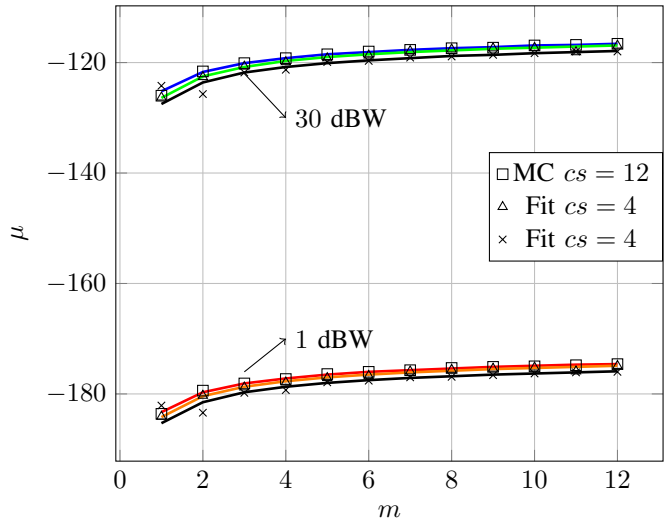
\begin{figure}
    \centering
    \begin{tikzpicture}
        \begin{axis}[
            xlabel={$m$},
            ylabel={$\mu$},
            legend style={
                at={(1,0.5)},
                anchor=mid east,
                },
            grid=both,
            width=0.5\textwidth,
        ]

        \addplot[
            color=black,
            only marks,
            mark=square,
        ] table [x=ns, y=y1, col sep=space] {GEV_Fitting_mu_30dBW.dat};

        \addplot[
            color=black,
            only marks,
            mark=triangle,
        ] table [x=ns, y=y2, col sep=space] {GEV_Fitting_mu_30dBW.dat};

        \addplot[
            color=black,
            only marks,
            mark=x,
        ] table [x=ns, y=y3, col sep=space] {GEV_Fitting_mu_30dBW.dat};
        \addlegendentry{MC $cs=12$}

        \addplot[
            color=blue,
            solid,
            line width=1pt,
        ] table [x=x1, y=yFit1, col sep=space] {GEV_Fitting_mu_30dBW.dat};
        \addlegendentry{Fit $cs=4$}

        \addplot[
            color=green,
            solid,
            line width=1pt,
        ] table [x=x2, y=yFit2, col sep=space] {GEV_Fitting_mu_30dBW.dat};

        \addplot[
            color=black,
            solid,
            line width=1pt,
        ] table [x=x3, y=yFit3, col sep=space] {GEV_Fitting_mu_30dBW.dat};

        \addplot[
            color=black,
            only marks,
            mark=square,
        ] table [x=ns, y=y1, col sep=space] {GEV_Fitting_mu_1dBW.dat};

        \addplot[
            color=black,
            only marks,
            mark=triangle,
        ] table [x=ns, y=y2, col sep=space] {GEV_Fitting_mu_1dBW.dat};

        \addplot[
            color=black,
            only marks,
            mark=x,
        ] table [x=ns, y=y3, col sep=space] {GEV_Fitting_mu_1dBW.dat};

        \addplot[
            color=red,
            solid,
            line width=1pt,
        ] table [x=x1, y=yFit1, col sep=space] {GEV_Fitting_mu_1dBW.dat};
        \addlegendentry{Fit $cs=4$}

        \addplot[
            color=orange,
            solid,
            line width=1pt,
        ] table [x=x2, y=yFit2, col sep=space] {GEV_Fitting_mu_1dBW.dat};

        \addplot[
            color=black,
            solid,
            line width=1pt,
        ] table [x=x3, y=yFit3, col sep=space] {GEV_Fitting_mu_1dBW.dat};

    \draw[->] (axis cs:3,-176) -- (axis cs:4,-170) node[right] {$1~\text{dBW}$};
    \draw[->] (axis cs:3,-122) -- (axis cs:4,-130) node[right] {$30~\text{dBW}$};

        \end{axis}
    \end{tikzpicture}
    \caption{\ac{GEV} $\mu$ fitting for different \ac{OFDM} symbols, comb size and $P_{\text{TX}}$ parameters.}
    \label{fig:GEV_Fitting_mu}
\end{figure}
Figure \ref{fig:GEV_Fitting_mu} shows that the value of $\mu$ depends on the number of symbols $m$ and the transmitted power $P_{\text{TX}}$. It is modeled by $\mu(m,P_{\text{TX}})=\frac{a_{1}}{\sqrt{m}}+a_{2}P_{\text{TX}}+a_{3}$. Then, to extract the values for $a_1,a_2,a_3$ we apply a change of variable $u=\frac{1}{\sqrt{m}}$. The equation can be written in matrix form as $\mathbf{F}=\mathbf{X}\beta$ where:

\begin{equation}
    \mathbf{F}=
    \begin{pmatrix}
        \mu_{1} \\
        \mu_{2} \\
        \vdots  \\
        \mu_{z}
    \end{pmatrix}, \quad
    \mathbf{X} = 
    \begin{pmatrix}
        1 & P_{\text{TX 1}} & 1 \\
        \frac{1}{\sqrt{2}} & P_{\text{TX 1}} & 1 \\
        \vdots & \vdots & \vdots \\
        \frac{1}{\sqrt{12}} & P_{\text{TX 1}} & 1 \\
        1 & P_{\text{TX 2}} & 1 \\
        \frac{1}{\sqrt{2}} & P_{\text{TX 2}} & 1 \\
        \vdots & \vdots & \vdots \\
        \frac{1}{\sqrt{12}} & P_{\text{TX 2}} & 1 \\
        \vdots & \vdots & \vdots \\
        1 & P_{\text{TX n}} & 1 \\
        \frac{1}{\sqrt{2}} & P_{\text{TX n}} & 1 \\
        \vdots & \vdots & \vdots \\
        \frac{1}{\sqrt{12}} & P_{\text{TX n}} & 1 \\
    \end{pmatrix}, \quad
    \mathbf{\beta} = 
    \begin{pmatrix}
        a_1 \\
        a_2 \\
        a_3
    \end{pmatrix}
\end{equation}
and we solve for $\beta$ using \ac{LS} as $\beta=\left(\mathbf{X}^T\mathbf{X}\right)^{-1}\mathbf{X}^T\mathbf{\mu}$. 
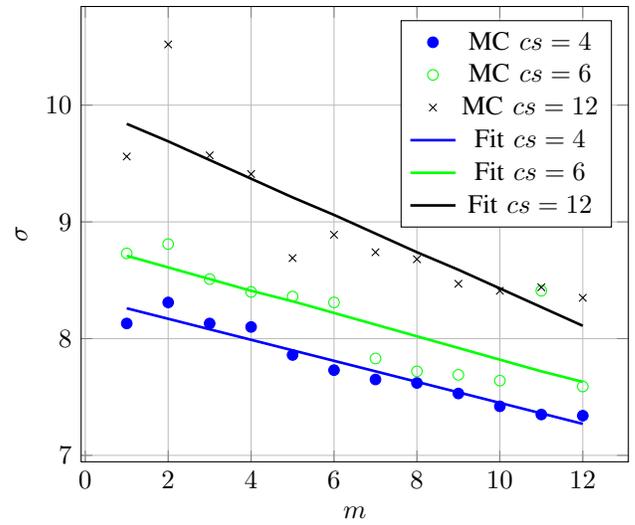
\begin{figure}
    \centering
    \begin{tikzpicture}
        \begin{axis}[
            xlabel={$m$},
            ylabel={$\sigma$},
            legend pos=north east,
            grid=both,
            width=0.5\textwidth,
        ]

        \addplot[
            color=blue,
            only marks,
            mark=*,
            mark options={solid},
        ] table [x=ns, y=y1, col sep=space] {GEV_Fitting_sigma.dat};
        \addlegendentry{MC $cs=4$}

        \addplot[
            color=green,
            only marks,
            mark=o,
            mark options={solid},
        ] table [x=ns, y=y2, col sep=space] {GEV_Fitting_sigma.dat};
        \addlegendentry{MC $cs=6$}

        \addplot[
            color=black,
            only marks,
            mark=x,
            mark options={solid},
        ] table [x=ns, y=y3, col sep=space] {GEV_Fitting_sigma.dat};
        \addlegendentry{MC $cs=12$}

        \addplot[
            color=blue,
            solid,
            line width=1pt,
        ] table [x=x1, y=yFit1, col sep=space] {GEV_Fitting_sigma.dat};
        \addlegendentry{Fit $cs=4$}

        \addplot[
            color=green,
            solid,
            line width=1pt,
        ] table [x=x2, y=yFit2, col sep=space] {GEV_Fitting_sigma.dat};
        \addlegendentry{Fit $cs=6$}

        \addplot[
            color=black,
            solid,
            line width=1pt,
        ] table [x=x3, y=yFit3, col sep=space] {GEV_Fitting_sigma.dat};
        \addlegendentry{Fit $cs=12$}

        \end{axis}
    \end{tikzpicture}
    \caption{\ac{GEV} $\sigma$ fitting for different \ac{OFDM} symbols and comb size parameters.}
    \label{fig:GEV_Fitting_sigma}
\end{figure}
\begin{figure}
    \centering
    \begin{tikzpicture}
        \begin{axis}[
            xlabel={$m$},
            ylabel={$k$},
            legend pos=south east,
            grid=both,
            width=0.45\textwidth,
        ]

        \addplot[
            color=blue,
            only marks,
            mark=*,
            mark options={solid},
        ] table [x=ns, y=y1, col sep=space] {GEV_Fitting_k.dat};
        \addlegendentry{MC $cs=4$}

        \addplot[
            color=green,
            only marks,
            mark=o,
            mark options={solid},
        ] table [x=ns, y=y2, col sep=space] {GEV_Fitting_k.dat};
        \addlegendentry{MC $cs=6$}

        \addplot[
            color=black,
            only marks,
            mark=x,
            mark options={solid},
        ] table [x=ns, y=y3, col sep=space] {GEV_Fitting_k.dat};
        \addlegendentry{MC $cs=12$}

        \addplot[
            color=blue,
            solid,
            line width=1pt,
        ] table [x=x1, y=yFit1, col sep=space] {GEV_Fitting_k.dat};
        \addlegendentry{Fit $cs=4$}

        \addplot[
            color=green,
            solid,
            line width=1pt,
        ] table [x=x2, y=yFit2, col sep=space] {GEV_Fitting_k.dat};
        \addlegendentry{Fit $cs=6$}

        \addplot[
            color=black,
            solid,
            line width=1pt,
        ] table [x=x3, y=yFit3, col sep=space] {GEV_Fitting_k.dat};
        \addlegendentry{Fit $cs=12$}

        \end{axis}
    \end{tikzpicture}
    \caption{\ac{GEV} $k$ fitting for different \ac{OFDM} symbols and comb size parameters.}
    \label{fig:GEV_Fitting_k}
\end{figure}
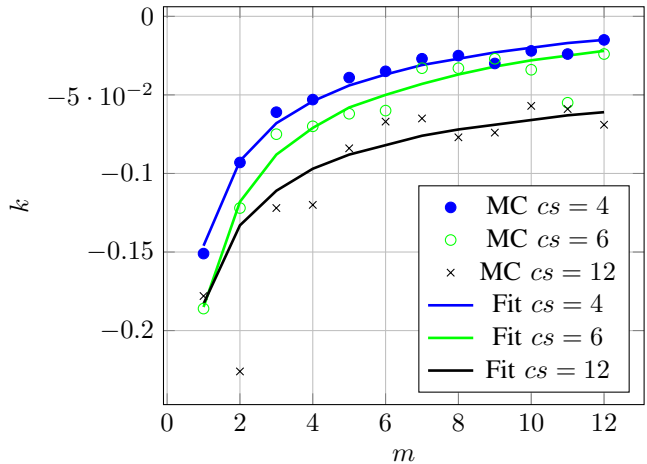
Then we model $\sigma$ from the results as a linear model $\sigma(m)=b_{1}m+b_{2}$ as seen in Figure \ref{fig:GEV_Fitting_sigma}. The last parameter we model is $k$, from the results they follow a curve $k(m)=\frac{c_{1}}{\sqrt{m}}+c_{2}$, as seen in Figure \ref{fig:GEV_Fitting_k}. Finally, table \ref{tab:GEV_model_params} presents the values for the different parameters to generate $[k(m),\mu(m,P_{\text{TX}}),\sigma(m)]$ for the \ac{GEV}.
\begin{table}
    \centering
    \begin{tabular}{|c|c|c|c|c|c|c|c|c|}
        \hline
        $cs$  & $a_1$ & $a_2$ & $a_3$ & $b_1$ & $b_2$ & $c_1$ & $c_2$ \\ \hline
        4   & 0.629&1.99 &-182 & -0.090 & 8.35 & -0.185 & 0.038 \\ \hline
        6   & 0.643&1.99 &-183 & -0.098 & 8.80 & -0.229 & 0.044 \\ \hline
        12  & 0.612&2.00 &-184 & -0.157 & 10 & -0.172 & -0.011  \\ \hline
    \end{tabular}
    \caption{Values of the parameters to generate $[k(m),\mu(m,P_{\text{TX}}),\sigma(m)]$ of the \ac{GEV} for the interference of the PRS.}
    \label{tab:GEV_model_params}
\end{table}
This model extraction can be used in future simulations to evaluate the interference power that a PRS pilot will generate in a typical scenario with reception from 4 satellites.
\section{Conclusions and Prospects for Future Research} \label{sec:Conclusions}

This study presents a model of the interference power in a \ac{NTN} scenario when using the \ac{PRS} as a signal for positioning. The model has been extracted empirically using a Monte Carlo approach as the dynamics of the \ac{LEO} scenario makes it unfeasible to have a direct mathematical formulation for interference modeling.
Two analytical tools, the \ac{CAF} and the \ac{DDM}, have been used in assessing the interference for a \ac{PRS} signal. These methodologies have paved the way for an open research question: the feasibility of designing a positioning system that multiplexes different transmissions while minimizing interference.
In summary, the findings of this study contribute significantly to the understanding of interference challenges in \ac{LEO} scenarios. They mark a step towards the integration of \ac{5G} \ac{NTN} as an autonomous system, independent of \ac{GNSS} receivers at the \ac{UE} end. The results highlight the need for advanced multiplexing strategies to manage multiple satellite signals within the same \ac{BWP}, thereby mitigating \ac{SINR} degradation. 
Future research includes applying a precompensation to delay and Doppler within the beam coverage area, including the \ac{ICI} and \ac{ISI} model during the \ac{DDM} computation to reduce their effects, or exploring alternative waveform designs such as \ac{OTFS}, which could leverage their robustness in high Doppler and delay environments. Another avenue for enhancing current results involves the implementation of a \ac{SIC} algorithm, which holds promise for improving the \ac{SINR} of the signal \cite{miridakis_survey_2013}, or exploit the sparcity of the \ac{DDM} matrix by using a \ac{OMP} algorithm.
\appendices
\section*{Appendix. Acronyms}
This paper makes use of an extensive number of acronyms, and to help the reader, the following list shows all of them:
\begin{acronym} [DL-OTDoA]
\acro{3G}{third generation}
\acro{4G}{fourth generation}
\acro{5G}{fifth generation}
\acro{6G}{sixth generation}
\acro{3GPP}{3rd Generation Partnership Project}
\acro{AI}{artificial intelligence}
\acro{AF}{ambiguity function}
\acro{AMF}{access and mobility function}
\acro{AoA}{angle of arrival}
\acro{AR}{augmented reality}
\acro{AWGN}{additive white Gaussian noise}
\acro{B5G}{beyond 5th generation}
\acro{BER}{bit error rate}
\acro{BW}{bandwidth}
\acro{BWP}{bandwidth part}
\acro{BWPP}{bandwidth part for positioning}
\acro{CA}{cell averaging}
\acro{CAF}{cross ambiguity function}
\acro{CDF}{cumulative density function}
\acro{CIR}{channel impulse response}
\acro{CFAR}{constant false alarm rate}
\acro{CP}{cyclix prefix}
\acro{CP-OFDM}{cyplix prefix orthogonal frequency-division modulation}
\acro{CRLB}{Cramer-Rao lower bound}
\acro{CU}{centralised unit}
\acro{DDM}{delay/Doppler map}
\acro{DFT}{discrete Fourier transform}
\acro{DL-AoD}{downlink angle of departure}
\acro{DL-TDoA}{downlink time difference of arrival}
\acro{DL-OTDoA}{downlink observed time difference of arrival}
\acro{DM-RS}{demodulation reference signal}
\acro{DoC}{depth of coverage}
\acro{DU}{distributed unit}
\acro{DVB}{digital video broadcasting}
\acro{E-CID}{enhanced cell ID}
\acro{ECDF}{empirical cumulative density function}
\acro{ECEF}{Earth centered Earth fixed}
\acro{E-SMLC}{evolved serving mobile location center}
\acro{EIRP}{equivalent isotropic radiated power}
\acro{eNB}{evolved nodeB}
\acro{FCC}{Federal Communications Commission}
\acro{FFR}{fraction frequency reuse}
\acro{FoV}{field of view}
\acro{FR}{frequency regions}
\acro{FR1}{frequency region 1}
\acro{FR2}{frequency region 2}
\acro{FSPL}{free space path loss}
\acro{FSS}{fixed satellite service}
\acro{GDOP}{geometrical dilution of precision}
\acro{GEV}{Generalized extreme value}
\acro{GPS}{global positioning system}
\acro{gNB}{next generation base station}
\acro{GNSS}{global navigation satellite system}
\acro{GS}{ground station}
\acro{HAPS}{high-altitude platform systems}
\acro{HPA}{high power amplifier}
\acro{IDFT}{inverse discrete Fourier transform}
\acro{IFFT}{inverse fast Fourier transform}
\acro{IoT}{Internet of things}
\acro{IIoT}{industrial Internet of things}
\acro{ICI}{inter-carrier interference}
\acro{IMU}{inertial measurement unit}
\acro{INI}{inter-numerology interference}
\acro{ISI}{inter-symbol interference}
\acro{KPI}{key performance indicator}
\acro{KS}{Kolmogorov-Smirnov test}
\acro{LCS}{location-based services}
\acro{LEO}{low Earth orbit}
\acro{LMC}{location management component}
\acro{LMF}{location management function}
\acro{LMS}{land mobile service}
\acro{LOS}{line of sight}
\acro{LPP}{localization positioning protocol}
\acro{LPPa}{localization positioning protocol annex}
\acro{LS}{least squares}
\acro{LTE}{Long Term Evolution}
\acro{LTI}{linear time-invariant}
\acro{LTV}{linear time-variant}
\acro{MIMO}{multiple-input multiple-ouput}
\acro{ML}{machine learning}
\acro{MLE}{maximum likelihood estimator}
\acro{MNO}{mobile network operator}
\acro{MSS}{mobile satellite service}
\acro{Multi-RTT}{multicell round trip time}
\acro{NF}{network function}
\acro{NLOS}{non-line of sight}
\acro{NPRM}{notice of proposed rulemaking}
\acro{NR}{new radio}
\acro{NTN}{non-terrestrial network}
\acro{OFDM}{orthogonal frequency-division multiplexing}
\acro{OMP}{orthogonal matching pursuit}
\acro{OTDoA}{observed time differential of arrival}
\acro{OTFS}{orthogonal time frequency space modulation}
\acro{PCS}{personal communication services}
\acro{PDF}{probability density function}
\acro{PNT}{positioning, navigation, and timing}
\acro{POD}{precise orbit determination}
\acro{PPP}{precise point positioning}
\acro{PRF}{pulse repetition frequency}
\acro{PRN}{pseudo-random noise}
\acro{PRS}{positioning reference signal}
\acro{PSS}{primary synchronization signal}
\acro{PVT}{position, velocity, and time}
\acro{QPSK}{quadrature phase-shift keying}
\acro{RAN}{radio access network}
\acro{RAT}{radio access technology}
\acro{RB}{resource block}
\acro{RE}{resource element}
\acro{RG}{resource grid}
\acro{RedCap}{reduced capacity}
\acro{RMSE}{root mean square error}
\acro{RTK}{real time kinematics}
\acro{RRC}{radio resource control}
\acro{SBAS}{satellite based augmentation system}
\acro{SCS}{subcarrier spacing}
\acro{SFN}{single frequency network}
\acro{SIB}{system information block}
\acro{SIC}{succesive interference cancellation}
\acro{SINR}{signal to interference plus noise ratio}
\acro{SLA}{service level agreement}
\acro{SNO}{satellite network operator}
\acro{SNR}{signal to noise ratio}
\acro{SoO}{signal of opportunity}
\acro{SoP}{signal of opportunity}
\acro{SRS}{sounding reference signal}
\acro{SSB}{synchronization system block}
\acro{SSP}{subsatellite point}
\acro{SSS}{secondary synchronization signal}
\acro{TA}{timing advance}
\acro{TDE}{time-domain equalizer}
\acro{TDL}{tapped delay line}
\acro{TN}{terrestrial network}
\acro{ToA}{time of arrival}
\acro{ToF}{time of flight}
\acro{TS}{technical specification}
\acro{TR}{technical report}
\acro{TTFF}{time to first fix}
\acro{UAV}{unmanned aerial vehicle}
\acro{UE}{user equipment}
\acro{UL-AoA}{uplink angle of arrival}
\acro{UL-TDoA}{uplink time difference of arrival}
\acro{VR}{virtual reality}
\acro{WI}{work item}
\acro{WLAN}{wireless local area network}
\acro{WLS}{weighted least squares}
\acro{WSS}{wide-sense stationary}
\acro{WSSUS}{wide-sense stationary uncorrelated scattering}
\end{acronym}

\section*{Acknowledgement}
\textbf{Acknowledgments:}
The simulations presented in this paper were carried out using the HPC facilities of the University of Luxembourg~\cite{VCPKVO_HPCCT22} (see \texttt{\href{http://hpc.uni.lu}{hpc.uni.lu}}).

The work presented in this paper was carried out in the framework of SatNEx V WI 3.3: LEO-PNT project, funded by ESA. contract no. 4000130962/20/NL/NL/FE. The authors thank Dr. Nader Alagha from ESA for his constructive discussions. The views expressed in this paper can in no way be taken to reflect the official opinion of the European Space Agency.
\bibliographystyle{IEEEtran}
\bibliography{main}

\begin{thebibliography}{10}
\providecommand{\url}[1]{#1}
\csname url@samestyle\endcsname
\providecommand{\newblock}{\relax}
\providecommand{\bibinfo}[2]{#2}
\providecommand{\BIBentrySTDinterwordspacing}{\spaceskip=0pt\relax}
\providecommand{\BIBentryALTinterwordstretchfactor}{4}
\providecommand{\BIBentryALTinterwordspacing}{\spaceskip=\fontdimen2\font plus
\BIBentryALTinterwordstretchfactor\fontdimen3\font minus \fontdimen4\font\relax}
\providecommand{\BIBforeignlanguage}[2]{{%
\expandafter\ifx\csname l@#1\endcsname\relax
\typeout{** WARNING: IEEEtran.bst: No hyphenation pattern has been}%
\typeout{** loaded for the language `#1'. Using the pattern for}%
\typeout{** the default language instead.}%
\else
\language=\csname l@#1\endcsname
\fi
#2}}
\providecommand{\BIBdecl}{\relax}
\BIBdecl

\bibitem{el_jaafari_introduction_2023}
\BIBentryALTinterwordspacing
M.~El~Jaafari, N.~Chuberre, S.~Anjuere, and L.~Combelles, ``\BIBforeignlanguage{en}{Introduction to the {3GPP}-defined {NTN} standard: {A} comprehensive view on the {3GPP} work on {NTN}},'' \emph{\BIBforeignlanguage{en}{International Journal of Satellite Communications and Networking}}, vol.~41, no.~3, pp. 220--238, 2023, \_eprint: https://onlinelibrary.wiley.com/doi/pdf/10.1002/sat.1471. [Online]. Available: \url{https://onlinelibrary.wiley.com/doi/abs/10.1002/sat.1471}
\BIBentrySTDinterwordspacing

\bibitem{noauthor_study_2019}
\BIBentryALTinterwordspacing
``Study on {NR} positioning support,'' 3GPP, Technical {Report} ({TR}) 38.855, 2019. [Online]. Available: \url{https://portal.3gpp.org/desktopmodules/Specifications/SpecificationDetails.aspx?specificationId=3501}
\BIBentrySTDinterwordspacing

\bibitem{muursepp_performance_2021}
\BIBentryALTinterwordspacing
I.~Muursepp, M.~Kulmar, O.~Elghary, M.~M. Alam, T.~Chen, S.~Horsmanheimo, and J.~Scholliers, ``\BIBforeignlanguage{en}{Performance {Evaluation} of {5G}-{NR} {Positioning} {Accuracy} {Using} {Time} {Difference} of {Arrival} {Method}},'' in \emph{\BIBforeignlanguage{en}{2021 {IEEE} {International} {Mediterranean} {Conference} on {Communications} and {Networking} ({MeditCom})}}.\hskip 1em plus 0.5em minus 0.4em\relax Athens, Greece: IEEE, Sep. 2021, pp. 494--499. [Online]. Available: \url{https://ieeexplore.ieee.org/document/9647652/}
\BIBentrySTDinterwordspacing

\bibitem{panchetti_performance_2013}
\BIBentryALTinterwordspacing
M.~Panchetti, C.~Carbonelli, M.~Horvat, and M.~Luise, ``\BIBforeignlanguage{en}{Performance analysis of {PRS}-based synchronization algorithms for {LTE} positioning applications},'' in \emph{\BIBforeignlanguage{en}{2013 10th {Workshop} on {Positioning}, {Navigation} and {Communication} ({WPNC})}}.\hskip 1em plus 0.5em minus 0.4em\relax Dresden: IEEE, Mar. 2013, pp. 1--6. [Online]. Available: \url{http://ieeexplore.ieee.org/document/6533292/}
\BIBentrySTDinterwordspacing

\bibitem{rinaldi_non-terrestrial_2020}
F.~Rinaldi, H.-L. Maattanen, J.~Torsner, S.~Pizzi, S.~Andreev, A.~Iera, Y.~Koucheryavy, and G.~Araniti, ``Non-{Terrestrial} {Networks} in {5G} \& {Beyond}: {A} {Survey},'' \emph{IEEE Access}, vol.~8, pp. 165\,178--165\,200, 2020, conference Name: IEEE Access.

\bibitem{trevlakis_localization_2023}
\BIBentryALTinterwordspacing
S.~E. Trevlakis, A.-A.~A. Boulogeorgos, D.~Pliatsios, J.~Querol, K.~Ntontin, P.~Sarigiannidis, S.~Chatzinotas, and M.~Di~Renzo, ``Localization as a {Key} {Enabler} of {6G} {Wireless} {Systems}: {A} {Comprehensive} {Survey} and an {Outlook},'' \emph{IEEE Open Journal of the Communications Society}, vol.~4, pp. 2733--2801, 2023, conference Name: IEEE Open Journal of the Communications Society. [Online]. Available: \url{https://ieeexplore.ieee.org/document/10287134/references#references}
\BIBentrySTDinterwordspacing

\bibitem{dureppagari_ntn-based_2023}
\BIBentryALTinterwordspacing
H.~K. Dureppagari, C.~Saha, H.~S. Dhillon, and R.~M. Buehrer, ``\BIBforeignlanguage{en}{{NTN}-based {6G} {Localization}: {Vision}, {Role} of {LEOs}, and {Open} {Problems}},'' Sep. 2023, arXiv:2305.12259 [cs, eess, math]. [Online]. Available: \url{http://arxiv.org/abs/2305.12259}
\BIBentrySTDinterwordspacing

\bibitem{wei_5g_2023}
Z.~Wei, Y.~Wang, L.~Ma, S.~Yang, Z.~Feng, C.~Pan, Q.~Zhang, Y.~Wang, H.~Wu, and P.~Zhang, ``{5G} {PRS}-{Based} {Sensing}: {A} {Sensing} {Reference} {Signal} {Approach} for {Joint} {Sensing} and {Communication} {System},'' \emph{IEEE Transactions on Vehicular Technology}, vol.~72, no.~3, pp. 3250--3263, Mar. 2023, conference Name: IEEE Transactions on Vehicular Technology.

\bibitem{martins_intersymbol_2019}
W.~A. Martins, F.~Cruz–Roldán, M.~Moonen, and P.~S. Ramirez~Diniz, ``Intersymbol and {Intercarrier} {Interference} in {OFDM} {Transmissions} {Through} {Highly} {Dispersive} {Channels},'' in \emph{2019 27th {European} {Signal} {Processing} {Conference} ({EUSIPCO})}, Sep. 2019, pp. 1--5, iSSN: 2076-1465.

\bibitem{cruz-roldan_intersymbol_2020}
\BIBentryALTinterwordspacing
F.~Cruz-Roldán, W.~A. Martins, F.~G. G., M.~Moonen, and P.~S.~R. Diniz, ``\BIBforeignlanguage{en}{Intersymbol and {Intercarrier} {Interference} in {OFDM} {Systems}: {Unified} {Formulation} and {Analysis}},'' Dec. 2020, arXiv:2012.04527 [eess]. [Online]. Available: \url{http://arxiv.org/abs/2012.04527}
\BIBentrySTDinterwordspacing

\bibitem{nemati_low_2018}
M.~Nemati and H.~Arslan, ``Low {ICI} {Symbol} {Boundary} {Alignment} for {5G} {Numerology} {Design},'' \emph{IEEE Access}, vol.~6, pp. 2356--2366, 2018, conference Name: IEEE Access.

\bibitem{marijanovic_multiplexing_2020}
L.~Marijanović, S.~Schwarz, and M.~Rupp, ``Multiplexing {Services} in {5G} and {Beyond}: {Optimal} {Resource} {Allocation} {Based} on {Mixed} {Numerology} and {Mini}-{Slots},'' \emph{IEEE Access}, vol.~8, pp. 209\,537--209\,555, 2020, conference Name: IEEE Access.

\bibitem{kihero_inter-numerology_2019}
\BIBentryALTinterwordspacing
A.~B. Kihero, M.~S.~J. Solaija, and H.~Arslan, ``\BIBforeignlanguage{en}{Inter-{Numerology} {Interference} for {Beyond} {5G}},'' \emph{\BIBforeignlanguage{en}{IEEE Access}}, vol.~7, pp. 146\,512--146\,523, 2019. [Online]. Available: \url{https://ieeexplore.ieee.org/document/8861343/}
\BIBentrySTDinterwordspacing

\bibitem{khan_novel_2019}
\BIBentryALTinterwordspacing
S.~A. Khan, A.~Kavak, s.~Aldirmaz~Çolak, and K.~Küçük, ``A {Novel} {Fractional} {Frequency} {Reuse} {Scheme} for {Interference} {Management} in {LTE}-{A} {HetNets},'' \emph{IEEE Access}, vol.~7, pp. 109\,662--109\,672, 2019, conference Name: IEEE Access. [Online]. Available: \url{https://ieeexplore.ieee.org/document/8790698}
\BIBentrySTDinterwordspacing

\bibitem{venkatesan_network_2007}
\BIBentryALTinterwordspacing
S.~Venkatesan, A.~Lozano, and R.~Valenzuela, ``Network {MIMO}: {Overcoming} {Intercell} {Interference} in {Indoor} {Wireless} {Systems},'' in \emph{2007 {Conference} {Record} of the {Forty}-{First} {Asilomar} {Conference} on {Signals}, {Systems} and {Computers}}, Nov. 2007, pp. 83--87, iSSN: 1058-6393. [Online]. Available: \url{https://ieeexplore.ieee.org/document/4487170}
\BIBentrySTDinterwordspacing

\bibitem{trabelsi_interference_2024}
\BIBentryALTinterwordspacing
N.~Trabelsi, L.~C. Fourati, and C.~S. Chen, ``Interference {Management} in {5G} and {Beyond} {Networks},'' Jan. 2024, arXiv:2401.01608 [cs, eess, math]. [Online]. Available: \url{http://arxiv.org/abs/2401.01608}
\BIBentrySTDinterwordspacing

\bibitem{zhen_energy-efficient_2021}
\BIBentryALTinterwordspacing
L.~Zhen, A.~K. Bashir, K.~Yu, Y.~D. Al-Otaibi, C.~H. Foh, and P.~Xiao, ``Energy-{Efficient} {Random} {Access} for {LEO} {Satellite}-{Assisted} {6G} {Internet} of {Remote} {Things},'' \emph{IEEE Internet of Things Journal}, vol.~8, no.~7, pp. 5114--5128, Apr. 2021, conference Name: IEEE Internet of Things Journal. [Online]. Available: \url{https://ieeexplore.ieee.org/document/9222142}
\BIBentrySTDinterwordspacing

\bibitem{lin_positioning_2017}
\BIBentryALTinterwordspacing
X.~Lin, J.~Bergman, F.~Gunnarsson, O.~Liberg, S.~M. Razavi, H.~S. Razaghi, H.~Rydn, and Y.~Sui, ``Positioning for the {Internet} of {Things}: {A} {3GPP} {Perspective},'' \emph{IEEE Communications Magazine}, vol.~55, no.~12, pp. 179--185, Dec. 2017, conference Name: IEEE Communications Magazine. [Online]. Available: \url{https://ieeexplore.ieee.org/document/8030544}
\BIBentrySTDinterwordspacing

\bibitem{sixin_doppler_2022}
\BIBentryALTinterwordspacing
W.~Sixin, T.~Xiaomei, L.~Xiaohui, F.~Wang, and Z.~Zhuang, ``\BIBforeignlanguage{en}{Doppler frequency-code phase division multiple access technique for {LEO} navigation signals},'' \emph{\BIBforeignlanguage{en}{GPS Solutions}}, vol.~26, no.~3, p.~98, Jun. 2022. [Online]. Available: \url{https://doi.org/10.1007/s10291-022-01283-7}
\BIBentrySTDinterwordspacing

\bibitem{dwivedi_positioning_2021}
\BIBentryALTinterwordspacing
S.~Dwivedi, R.~Shreevastav, F.~Munier, J.~Nygren, I.~Siomina, Y.~Lyazidi, D.~Shrestha, G.~Lindmark, P.~Ernstrom, E.~Stare, S.~M. Razavi, S.~Muruganathan, G.~Masini, A.~Busin, and F.~Gunnarsson, ``Positioning in {5G} {Networks},'' \emph{IEEE Communications Magazine}, vol.~59, no.~11, pp. 38--44, Nov. 2021, conference Name: IEEE Communications Magazine. [Online]. Available: \url{https://ieeexplore.ieee.org/document/9665436}
\BIBentrySTDinterwordspacing

\bibitem{honnaiah_demand-driven_2023}
\BIBentryALTinterwordspacing
P.~J. Honnaiah, E.~Lagunas, S.~Chatzinotas, and J.~Krause, ``Demand-{Driven} {Beam} {Densification} in {Multibeam} {Satellite} {Communication} {Systems},'' \emph{IEEE Transactions on Aerospace and Electronic Systems}, vol.~59, no.~5, pp. 6534--6554, Oct. 2023, conference Name: IEEE Transactions on Aerospace and Electronic Systems. [Online]. Available: \url{https://ieeexplore.ieee.org/document/10130323}
\BIBentrySTDinterwordspacing

\bibitem{hong_delay-doppler_2022}
Y.~Hong, T.~Thaj, and E.~Viterbo, \emph{\BIBforeignlanguage{eng}{Delay-{Doppler} communications principles and applications}}.\hskip 1em plus 0.5em minus 0.4em\relax London: Academic Press, an imprint of Elsevier, 2022, oCLC: 1297827135.

\bibitem{klobuchar_ionospheric_1987}
\BIBentryALTinterwordspacing
J.~A. Klobuchar, ``Ionospheric {Time}-{Delay} {Algorithm} for {Single}-{Frequency} {GPS} {Users},'' \emph{IEEE Transactions on Aerospace and Electronic Systems}, vol. AES-23, no.~3, pp. 325--331, May 1987, conference Name: IEEE Transactions on Aerospace and Electronic Systems. [Online]. Available: \url{https://ieeexplore.ieee.org/document/4104345}
\BIBentrySTDinterwordspacing

\bibitem{di_giovanni_analytical_1990}
\BIBentryALTinterwordspacing
G.~Di~Giovanni and S.~M. Radicella, ``An analytical model of the electron density profile in the ionosphere,'' \emph{Advances in Space Research}, vol.~10, no.~11, pp. 27--30, Jan. 1990. [Online]. Available: \url{https://www.sciencedirect.com/science/article/pii/027311779090301F}
\BIBentrySTDinterwordspacing

\bibitem{sanz_subirana_gnss_2013}
J.~Sanz~Subirana, J.~M. Juan~Zornoza, and M.~Hernández-Pajares, \emph{\BIBforeignlanguage{English}{{GNSS} {DATA} {Processing}. {Fundamentals} and algorithms. {Vol}. 1}}.\hskip 1em plus 0.5em minus 0.4em\relax Noordwijk: ESA Communications, 2013, oCLC: 922681096.

\bibitem{qggt-xr49-24}
\BIBentryALTinterwordspacing
A.~Gonzalez-Garrido, ``Starlink satellite passes,'' 2024, tex.entrytype: data. [Online]. Available: \url{https://dx.doi.org/10.21227/qggt-xr49}
\BIBentrySTDinterwordspacing

\bibitem{baeza_overview_2022}
\BIBentryALTinterwordspacing
V.~M. Baeza, E.~Lagunas, H.~Al-Hraishawi, and S.~Chatzinotas, ``An {Overview} of {Channel} {Models} for {NGSO} {Satellites},'' in \emph{2022 {IEEE} 96th {Vehicular} {Technology} {Conference} ({VTC2022}-{Fall})}, Sep. 2022, pp. 1--6, iSSN: 2577-2465. [Online]. Available: \url{https://ieeexplore.ieee.org/document/10012693}
\BIBentrySTDinterwordspacing

\bibitem{3gpp_nr_2024}
\BIBentryALTinterwordspacing
3GPP, ``{NR}; {Physical} channels and modulation,'' 3GPP, Technical {Specification} ({TS}) 38.211, 2024. [Online]. Available: \url{https://portal.3gpp.org/desktopmodules/Specifications/SpecificationDetails.aspx?specificationId=3213}
\BIBentrySTDinterwordspacing

\bibitem{querol_snr_2016}
J.~Querol, A.~Alonso-Arroyo, R.~Onrubia, D.~Pascual, H.~Park, and A.~Camps, ``{SNR} {Degradation} in {GNSS}-{R} {Measurements} {Under} the {Effects} of {Radio}-{Frequency} {Interference},'' \emph{IEEE Journal of Selected Topics in Applied Earth Observations and Remote Sensing}, vol.~9, no.~10, pp. 4865--4878, Oct. 2016, conference Name: IEEE Journal of Selected Topics in Applied Earth Observations and Remote Sensing.

\bibitem{miridakis_survey_2013}
N.~I. Miridakis and D.~D. Vergados, ``A {Survey} on the {Successive} {Interference} {Cancellation} {Performance} for {Single}-{Antenna} and {Multiple}-{Antenna} {OFDM} {Systems},'' \emph{IEEE Communications Surveys \& Tutorials}, vol.~15, no.~1, pp. 312--335, 2013, conference Name: IEEE Communications Surveys \& Tutorials.

\bibitem{VCPKVO_HPCCT22}
S.~Varrette, H.~Cartiaux, S.~Peter, E.~Kieffer, T.~Valette, and A.~Olloh, ``Management of an academic {HPC} \& research computing facility: {The} {ULHPC} experience 2.0,'' in \emph{Proc. of the 6th {ACM} high performance computing and cluster technologies conf. ({HPCCT} 2022)}.\hskip 1em plus 0.5em minus 0.4em\relax Fuzhou, China: Association for Computing Machinery (ACM), Jul. 2022.

\end{thebibliography}
\begin{IEEEbiography}
[{\includegraphics[width=1in,height=1.25in,clip,keepaspectratio]{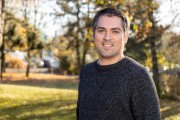}}]
{Alejandro Gonzalez-Garrido} ~ PhD Student at SIGCOM group from SnT, (University of Luxembourg) in hybrid GNSS and 5G PNT systems using Non-Terrestrial Networks. Integrated degree plus M.Sc. in Telecommunication Engineering obtained in 2015. With professional experience in the timing and synchronization industry, satellite design, and network operations. 
\end{IEEEbiography}

\begin{IEEEbiography}
[{\includegraphics[width=1in,height=1.25in,clip,keepaspectratio]{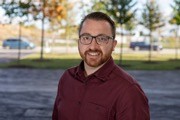}}]
{Jorge Querol} ~ received his Ph.D degree in telecommunication engineering, from the Polytechnic University of Catalonia (UPC-BarcelonaTech), Barcelona (Spain), in 2018. His research interests include Software Defined Radios (SDR), real-time signal processing, satellite communications, satellite navigation and remote sensing. Jorge joined the Signal Processing and Satellite Communications group, SIGCOM, headed by Prof. Bj\"orn Ottersten and he will be working with Dr. Symeon Chatzinotas.
\end{IEEEbiography}

\begin{IEEEbiography}
[{\includegraphics[width=1in,height=1.25in,clip,keepaspectratio]{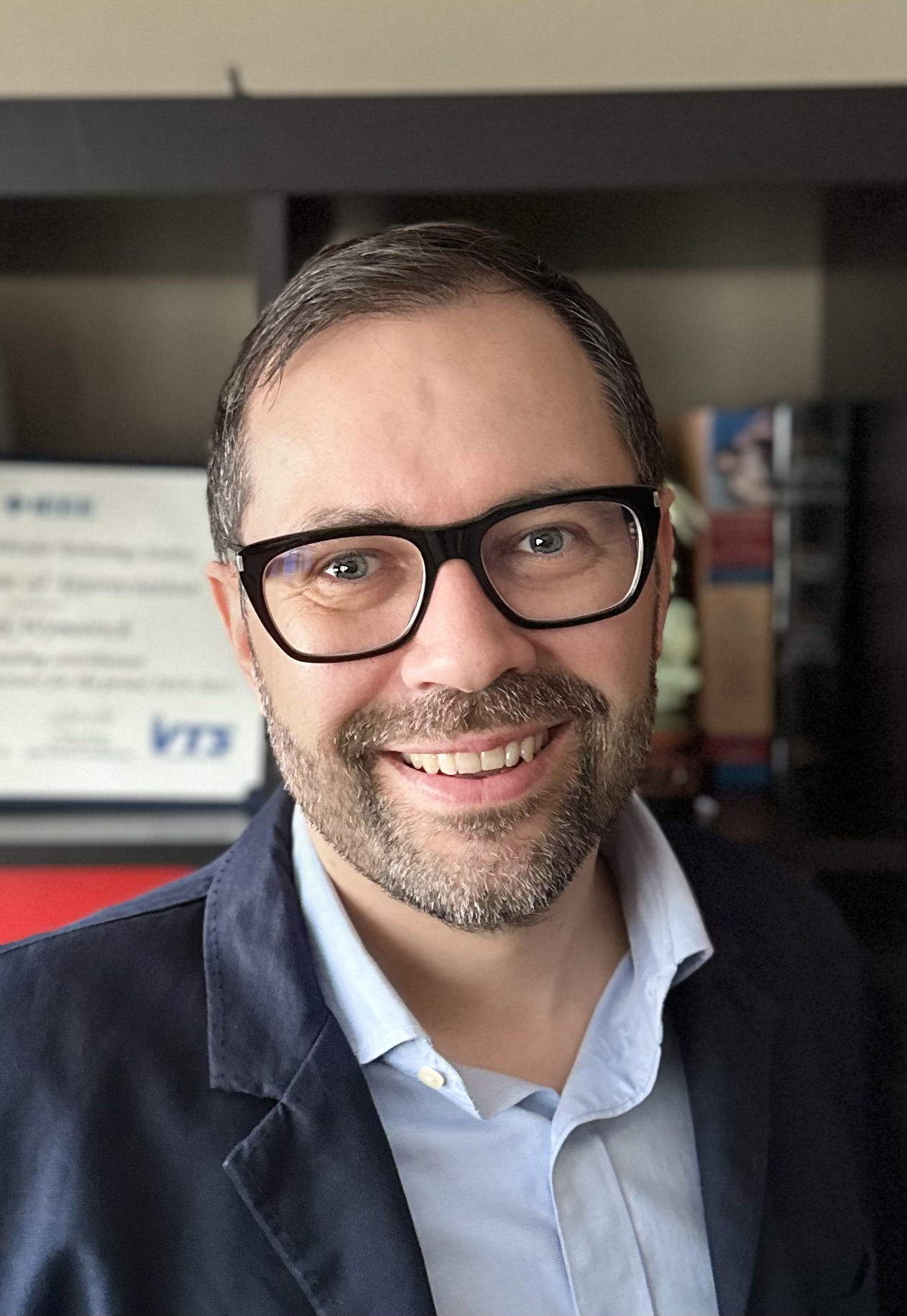}}]
{Henk Wymeersch}~ (S’01, M’05, SM’19, F’24) obtained the Ph.D. degree in Electrical Engineering/Applied Sciences in 2005 from Ghent University, Belgium. He is currently a Professor of Communication Systems with the Department of Electrical Engineering at Chalmers University of Technology, Sweden. He is Senior Member of the IEEE Signal Processing Magazine Editorial Board. During 2019-2021, he was an IEEE Distinguished Lecturer with the Vehicular Technology Society. His current research interests include the convergence of communication and sensing, in a 5G and Beyond 5G context.
\end{IEEEbiography}

\begin{IEEEbiography}
[{\includegraphics[width=1in,height=1.25in,clip,keepaspectratio]{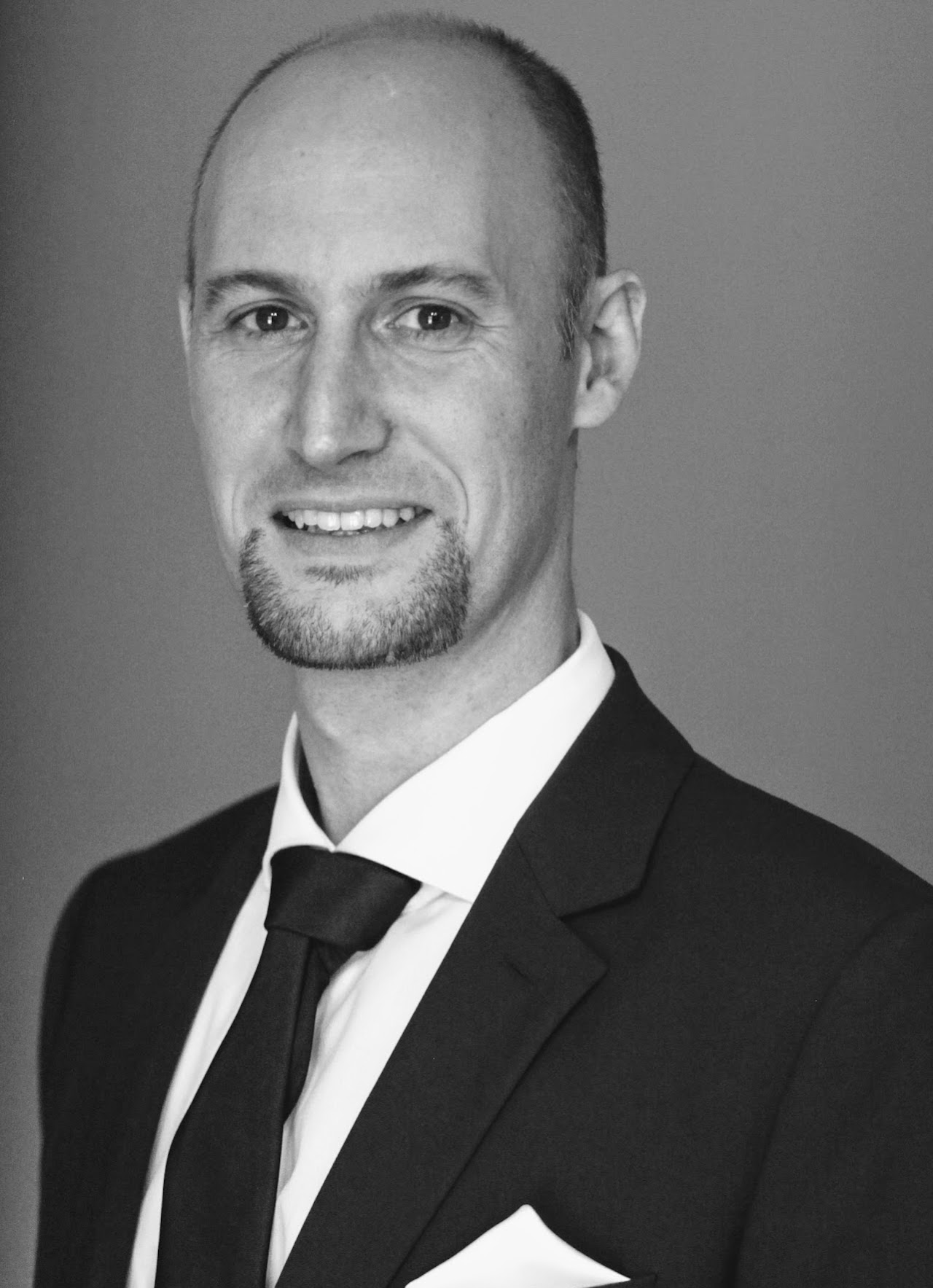}}]
{Symeon Chatzinotas}~ (MEng, MSc, PhD, F.IEEE) is currently Full Professor / Chief Scientist I and Head of the research group SIGCOM in the Interdisciplinary Centre for Security, Reliability and Trust, University of Luxembourg.

In the past, he has lectured as Visiting Professor at the University of Parma, Italy and contributed in numerous R\&D projects for the Institute of Informatics \& Telecommunications, National Center for Scientific Research ''Demokritos'' the Institute of Telematics and Informatics, Center of Research and Technology Hellas and Mobile Communications Research Group, Center of Communication Systems Research, University of Surrey.
He has received the M.Eng. in Telecommunications from Aristotle University of Thessaloniki, Greece and the M.Sc. and Ph.D. in Electronic Engineering from University of Surrey, UK in 2003, 2006 and 2009 respectively.
He has authored more than 700 technical papers in refereed international journals, conferences and scientific books and has received numerous awards and recognitions, including the IEEE Fellowship and an IEEE Distinguished Contributions Award. He is currently in the editorial board of the IEEE Transactions on Communications, IEEE Open Journal of Vehicular Technology and the International Journal of Satellite Communications and Networking.
\end{IEEEbiography}

\end{document}